\documentclass[runningheads]{llncs}
\usepackage{booktabs}
\usepackage{lineno}
\usepackage{tabularx}
\usepackage{multirow}
\usepackage{pdflscape}
\usepackage{setspace}
\usepackage[ruled,vlined,linesnumbered]{algorithm2e}
\usepackage{amsmath}
\usepackage{amssymb}
\usepackage{float}
\usepackage{bm}
\usepackage{mathtools}
\usepackage[table]{xcolor}
\usepackage{xspace}
\usepackage{multirow}
\usepackage{setspace}
\usepackage{tikz}
\usepackage{color}
\usepackage{fullpage}
\usepackage[bookmarks=false,colorlinks]{hyperref}
\usepackage[capitalize]{cleveref}
\crefname{problem}{problem}{problems}
\Crefname{problem}{Problem}{Problems}
\usepackage[labelformat=simple]{subcaption}

\usepackage{pgf,tikz,pgfplots}
\pgfplotsset{compat=1.15}
\usepackage{mathrsfs}
\usetikzlibrary{arrows}
\newtheorem{modl}{Model}
\newenvironment{model}{\begin{samepage}\begin{modl}}{\end{modl}\end{samepage}}

\renewcommand{\model}{\mathcal{M}}
\newcommand{\vars}{\mathcal{V}}
\newcommand{\cons}{\mathcal{C}}
\newcommand{\scon}{\mathcal{S}}
\newcommand{\rcon}{\mathcal{R}}

\newcommand{\repeatthanks}{\textsuperscript{\thefootnote}}

\usepackage{graphicx}

\title{A Safety Framework for Flow Decomposition Problems via Integer Linear Programming}
\author{Fernando H. C. Dias\inst{1,}\thanks{Shared first-author contribution}\orcidID{0000-0002-6398-919X} \and
Manuel C\'aceres\inst{1,}\repeatthanks\orcidID{0000-0003-0235-6951} \and
Lucia Williams\inst{2,}\repeatthanks\orcidID{0000-0003-3785-0247} \and
Brendan Mumey\inst{2,}\thanks{Shared last-author contribution}\orcidID{0000-0001-7151-2124} \and
Alexandru~I.~Tomescu\inst{1,}\repeatthanks\orcidID{0000-0002-5747-8350}}
\institute{Department of Computer Science, University of Helsinki, Finland \email{\{fernando.cunhadias,manuel.caceres,alexandru.tomescu\}@helsinki.fi}
\and
School of Computing, Montana State University, Bozeman, MT, USA
\email{\{lucia.williams,brendan.mumey\}@montana.edu}
}

\begin{document}

\maketitle

\begin{abstract}
Many important problems in Bioinformatics (e.g., assembly or multi-assembly) admit multiple solutions, while the final objective is to report only one. A common approach to deal with this uncertainty is finding \emph{safe} partial solutions (e.g., contigs) which are common to all solutions. Previous research on safety has focused on polynomially-time solvable problems, whereas many successful and natural models are NP-hard to solve, leaving a lack of ``safety tools" for such problems. We propose the first method for computing all safe solutions for an NP-hard problem, \emph{minimum flow decomposition}. We obtain our results by developing a “safety test” for paths based on a general Integer Linear Programming (ILP) formulation. Moreover, we provide implementations with practical optimizations aimed to reduce the total ILP time, the most efficient of these being based on a recursive group-testing procedure.\\
\textbf{Results:}
Experimental results on the transcriptome datasets of Shao and Kingsford (TCBB, 2017) show that all safe paths for minimum flow decompositions correctly recover up to 90\% of the full RNA transcripts, which is at least 25\% more than previously known safe paths, such as (C\'aceres et al. TCBB, 2021), (Zheng et al., RECOMB~2021), (Khan et al., RECOMB~2022, ESA~2022). Moreover, despite the NP-hardness of the problem,  we can report all safe paths for 99.8\% of the over 27,000 non-trivial graphs of this dataset in only 1.5 hours. Our results suggest that, on perfect data, there is less ambiguity than thought in the notoriously hard RNA assembly problem.\\ 
\textbf{Availability:} https://github.com/algbio/mfd-safety\\
\textbf{Contact:}  alexandru.tomescu@helsinki.fi\\

\end{abstract}

\newpage

\clearpage
\pagenumbering{arabic}

\section{Introduction}

In real-world scenarios where an unknown object needs to be discovered from the input data, we would like to formulate a computational problem loosely enough so that the unknown object is indeed a solution to the problem, but also tightly enough so that the problem does not admit many other solutions. However, this goal is difficult in practice, and indeed, various commonly used problem formulations in Bioinformatics still admit many solutions. While a naive approach is to just exhaustively enumerate all these solutions, a more practical approach is to report only those sub-solutions (or partial solutions) that are common to \emph{all} solutions to the problem.

In the graph theory community such sub-solutions have been called \emph{persistent}~\cite{Costa1994143,doi:10.1137/0603052}, and in the Bioinformatics community \emph{reliable}~\cite{vingron1990determination}, or more recently, \emph{safe}~\cite{tomescu2017safe}. The study of safe sub-solutions started in Bioinformatics in the 1990's \cite{vingron1990determination,Chao93,naor1994near} with those amino-acid pairs that are common to all optimal and suboptimal alignments of two protein sequences.

In the genome assembly community, the notion of \emph{contig}, namely a string that is guaranteed to appear in any possible assembly of the reads, is at the core of most genome assemblers. This approach originated in 1995 with the notion of unitigs~\cite{KM95} (non-branching paths in an assembly graph),
which were progressively~\cite{PTW01,BONNICI202123} generalized to paths made up of a prefix of nodes with in-degree one followed by nodes with out-degree one~\cite{MGMB07,jacksonthesis,kingsford2010assembly} (also called extended unitigs, or Y-to-V contigs).

Later, \cite{tomescu2017safe} formalized all such types of contigs as those \emph{safe} strings that appear in all solutions to a genome assembly problem formulation, expressed as a certain type of walk in a graph. \cite{cairo2019optimal,hydrostructure} proposed more efficient and unifying safety algorithms for several types of graph walks. \cite{rahman2022assembler} recently studied the safety of contigs produced by state-of-the-art genome assemblers on real data.

Analogous studies were recently made also for multi-assembly problems, where several related genomic sequences need to be assembled from a sample of mixed reads. \cite{Caceres:2022ve} studied safe paths that appear in all constrained path covers of a directed acyclic graph (DAG). Zheng, Ma and Kingsford studied the more practical setting of a network flow in a DAG by finding those paths that appear in any flow decomposition of the given network flow, under a probabilistic framework~\cite{ma2021exact}, or a combinatorial framework~\cite{findingranges}.\footnote{The problem AND-Quant from \cite{findingranges} actually handles a more general version of this problem.} \cite{khan2022safety} presented a simple characterization of safe paths appearing in any flow decomposition of a given acyclic network flow, leading to a more efficient algorithm than the one of~\cite{findingranges}, and further optimized by \cite{Khan:2022vi}.

\paragraph{\textbf{Motivation.}} Despite the significant progress in obtaining safe algorithms for a range of different applications, current safe algorithms are limited to problems where computing a solution itself is achievable in polynomial time. However, many natural problems are NP-hard, and safe algorithms for such problems are fully missing. Apart from the theoretical interest, usually such NP-hard problems correspond to restrictions of easier (polynomially-computable) problems, and thus by definition, also have longer safe sub-solutions.

As such, current safety algorithms miss data that could be reported as correct, just because they do not constrain the solution space enough. A major reason for this lack of progress is that if a problem is NP-hard, then its safety version is likely to be hard too. This phenomenon can be found both in classically studied NP-hard problems --- for example, computing nodes present in all maximum independent sets of an undirected graph is NP-hard~\cite{doi:10.1137/0603052} --- as well as in NP-hard problems studied for their application to Bioinformatics, as we discuss further in the appendix.

We introduce our results by focusing on the \emph{flow decomposition problem}. This is a classical model at the core of multi-assembly software for RNA transcripts \cite{li2011isolasso,li2011sparse,bernard2014efficient,tomescu2013novel} and viral quasi-species genomes \cite{baaijens2019full,baaijens2020strain,posada2021v,chen2018novo}, and also a standard problem with applications in other fields, such as networking \cite{mumey2015parity,hartman2012split,cohen2014effect,hong2013achieving} or transportation \cite{olsen2020study,Ohst:2015aa}. In its most basic optimization form, \emph{minimum flow decomposition (MFD)}, we are given a flow in a graph, and we need to decompose it into a minimum number of paths with associated weights, such that the superposition of these weighted paths gives the original flow. This is an NP-hard problem, even when restricted to DAGs~\cite{VATINLEN20081390,hartman2012split}. Various approaches have been proposed to tackle the problem, including fixed-parameter tractable algorithms~\cite{kloster2018practical}, approximation algorithms~\cite{mumey2015parity,DBLP:conf/esa/CaceresCG0MRTW22} and Integer Linear Programming formulations~\cite{dias2022fast,jumper}.

In Bioinformatics applications, reads or contigs originating from a mixed sample of genomic sequences with different abundances are aligned to a reference. A graph model, such as a splice graph or a variation graph, is built from these alignments. Read abundances assigned to the nodes and edges of this graph then correspond to a flow in case of perfect data. If this is not the case, the abundance values can either be minimally corrected to become a flow, or one can consider variations of the problem where e.g., the superposition of the weighted paths is closest (or within a certain range) to the edge abundances~\cite{tomescu2013novel,bernard2014efficient}.

Current safety algorithms for flow decompositions such as \cite{findingranges,khan2022safety,Khan:2022wo,Khan:2022vi} compute paths appearing in \emph{all} possible flow decompositions (of any size), even though decompositions of minimum size are assumed to better model the RNA assembly problem~\cite{kloster2018practical,shao2017theory,williams2021flow}. Even dropping the minimality constraint, but adding other simple constraints easily renders the problem NP-hard (see e.g.,~\cite{Williams:2022ws}), motivating further study of practical safe algorithms for NP-hard problems.

\paragraph{\textbf{Contributions.}} Integer Linear Programming (ILP) is a general and flexible method that has been successfully applied to solve NP-hard problems, including in Bioinformatics. In this paper, we consider graph problems whose solution consists of a set of \emph{paths} (i.e., not repeating nodes) that can be formulated in ILP. We introduce a technique that, given an ILP formulation of such a graph problem, can enhance it with additional variables and constraints in order to test the safety of a given set of paths. An obvious first application of this safety test is to use it with a single path in a straightforward avoid-and-test approach, using a standard two-pointer technique that has been used previously to find safe paths for flow decomposition. However, we find that a top-down recursive approach that uses the group-testing capability halves the number of computationally-intensive ILP calls, resulting in a 3x speedup over the straightforward approach.

Additionally, we prove that computing all the safe paths for MFDs is an intractable problem, confirming the above intuitive claim that if a problem is hard, then also its safety version is hard. We give this proof in the appendix by showing that the NP-hardness reduction for MFD by~\cite{hartman2012split} can be modified into a Turing reduction from the UNIQUE 3SAT problem.

On the dataset \cite{shao2017accurate} containing splice graphs from human, zebrafish and mouse transcriptomes, safe paths for MFDs (\emph{SafeMFD}) correctly recover up to 90\% of the full RNA transcripts while maintaining a 99\% precision, outperforming, by a wide margin (25\% increase), state-of-the-art safety approaches, such as extended unitigs~\cite{MGMB07,jacksonthesis,kingsford2010assembly}, safe paths for constrained path covers of the edges~\cite{Caceres:2022ve}, and safe paths for all flow decompositions~\cite{Khan:2022vi,khan2022safety,Khan:2022wo,findingranges}. On the harder dataset by \cite{Khan:2022wo}, \emph{SafeMFD} also dominates in a significant proportion of splice graphs (built from $t\le 15$ RNA transcripts), recovering more than 95\% of the full transcripts while maintaining a 98\% precision. For larger $t$, precision drastically drops (91\% precision in the entire dataset), suggesting that in more complex splice graphs smaller solutions are introduced as an artifact of the combinatorial nature of the splice graph, and the minimality condition~\cite{kloster2018practical,shao2017theory,williams2021flow} is thus incorrect in this domain.

\section{Methods}
\subsection{Preliminaries}
\label{sec:preliminaries}

\paragraph{\textbf{ILP models.}} In this paper we use ILP models as blackboxes, with as few assumptions as possible to further underline the generality of our approach. Let $\model(\vars,\cons)$ be an ILP model consisting of a set $\vars$ of variables and a set $\cons$ of constraints on these variables, built from an input graph $G = (V,E)$. We make only two assumptions on $\model$. First, that a solution to this model consists of a given number $k \geq 1$ of paths $P_1,\dots,P_k$ in $G$ (in this paper, paths do not repeat vertices). Second, we assume that the $k$ paths are modeled via binary edge variables $x_{uvi}$, for all $(u,v) \in E$ and for all $i \in \{1,\dots,k\}$. More specifically, for all $i \in \{1,\dots,k\}$, we require that the edges $(u,v) \in E$ for which the corresponding variable $x_{uvi}$ equals 1 induce a path in $G$. For example, if $G$ is a DAG, it is a standard fact (see e.g.,~\cite{taccari2016integer}) that a path from a given $s \in V$ to a given $t \in V$ (an \emph{$s$-$t$ path}) can be expressed with the following constraints:
\begin{equation}
\label{eqn:flow_conservation}
    \sum_{(u,v) \in E} x_{uvi} - \sum_{(v,u) \in E} x_{vui} =
    \begin{cases}
    0, & \text{if $v \in V \setminus \{s,t\}$}, \\
    1, & \text{if $v = t$}, \\
    -1, & \text{if $v = s$}.
    \end{cases}
\end{equation}

If $G$ is not a DAG, there are other types of constraints that can be added to the $x_{uvi}$ variables to ensure that they induce a path; see, for example, the many formulations in~\cite{taccari2016integer}. We will assume that such constraints are part of the set $\cons$ of constraints of $\model(\vars,\cons)$, but their exact formulation is immaterial for our approach. In fact, one could even add additional constraints to $\cons$ to further restrict the solution space. For example, some ILP models from \cite{dias2022fast,jumper} handle the case when the input also contains a set of paths (\emph{subpath constraints}) that must appear in at least one of the $k$ solution paths.

\begin{figure}
\allowdisplaybreaks
\centering
\begin{subfigure}[c]{0.45\textwidth}
\centering
\includegraphics[width=0.6\textwidth]{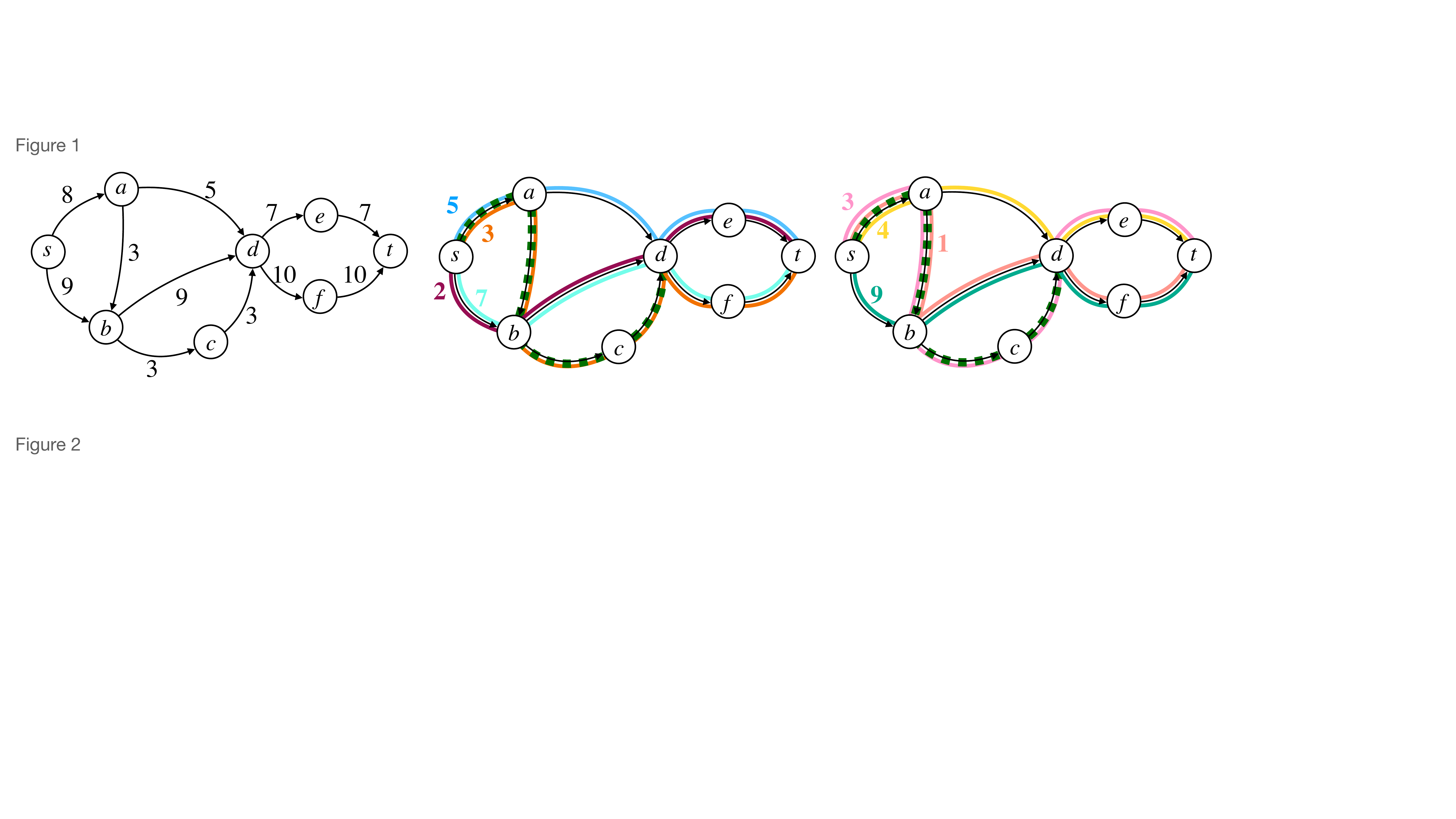}
\caption{A flow network with source $s$ and sink $t$.
\label{fig:flow-network-example}}
\end{subfigure}
\hspace{1cm}
\begin{subfigure}[c]{0.45\textwidth}
\centering
\includegraphics[width=0.6\textwidth]{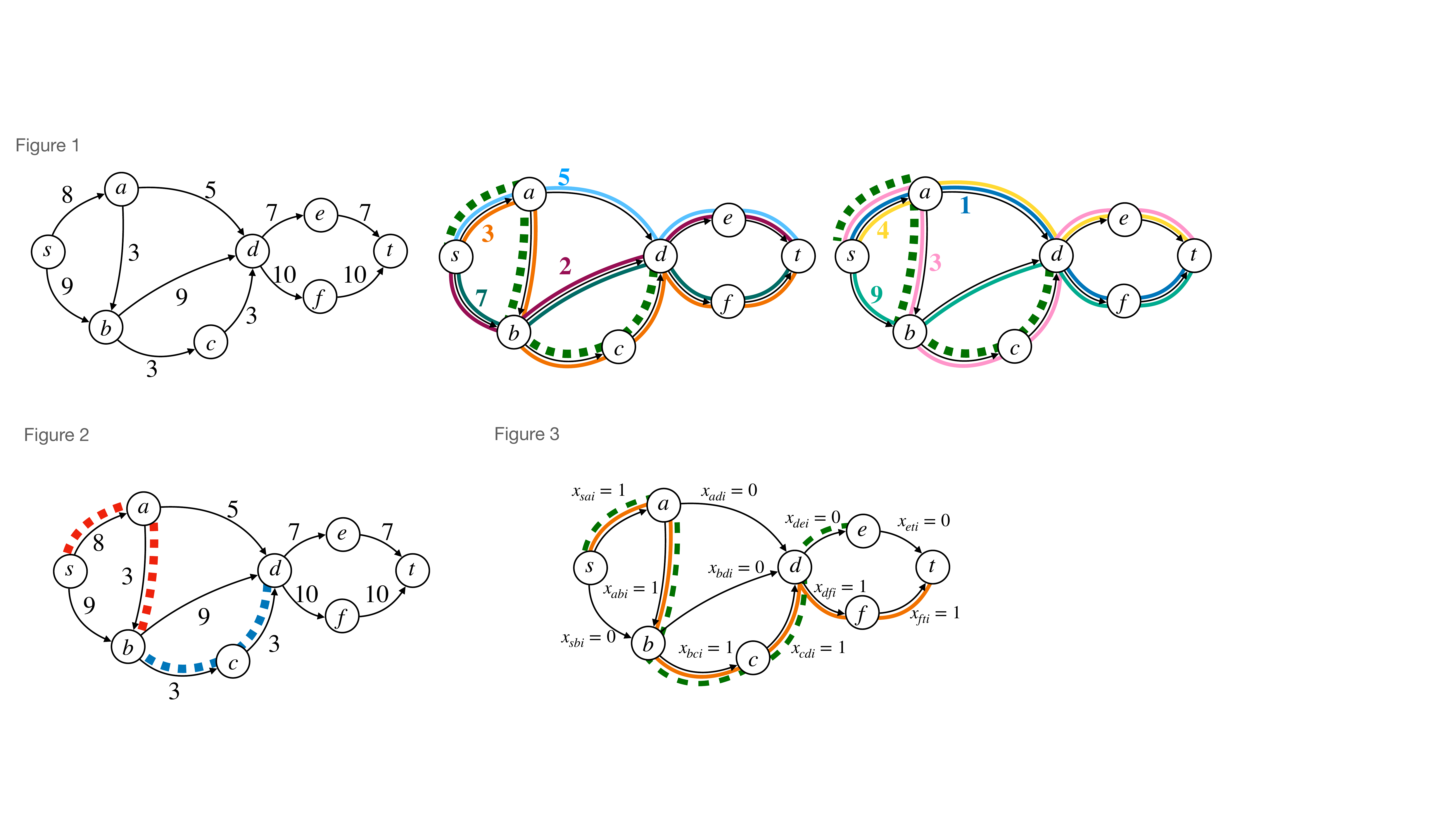}
\caption{An MFD into 4 paths of weights 5,3,7,2, respectively. The green dashed path is a subpath of the orange path. \label{fig:FD-example-1}}
\end{subfigure}
\begin{subfigure}[c]{0.45\textwidth}
\centering
\includegraphics[width=0.6\textwidth]{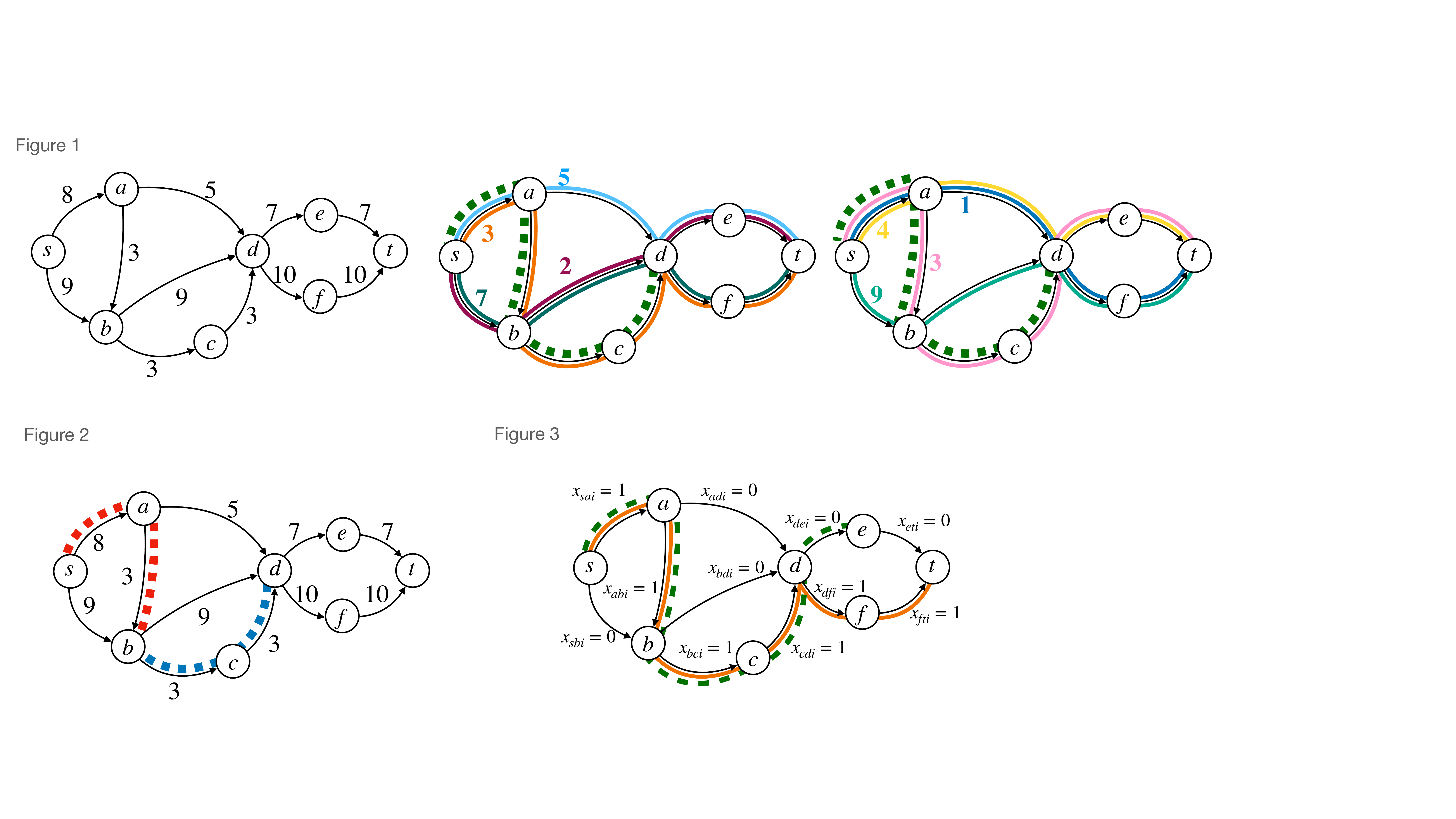}
\caption{An MFD into 4 paths of weights 3,4,1,9, respectively. The green dashed path is a subpath of the pink path.\label{fig:FD-example-2}}
\end{subfigure}
\hspace{1cm}
\begin{subfigure}[c]{0.45\textwidth}
\centering
\includegraphics[width=0.6\textwidth]{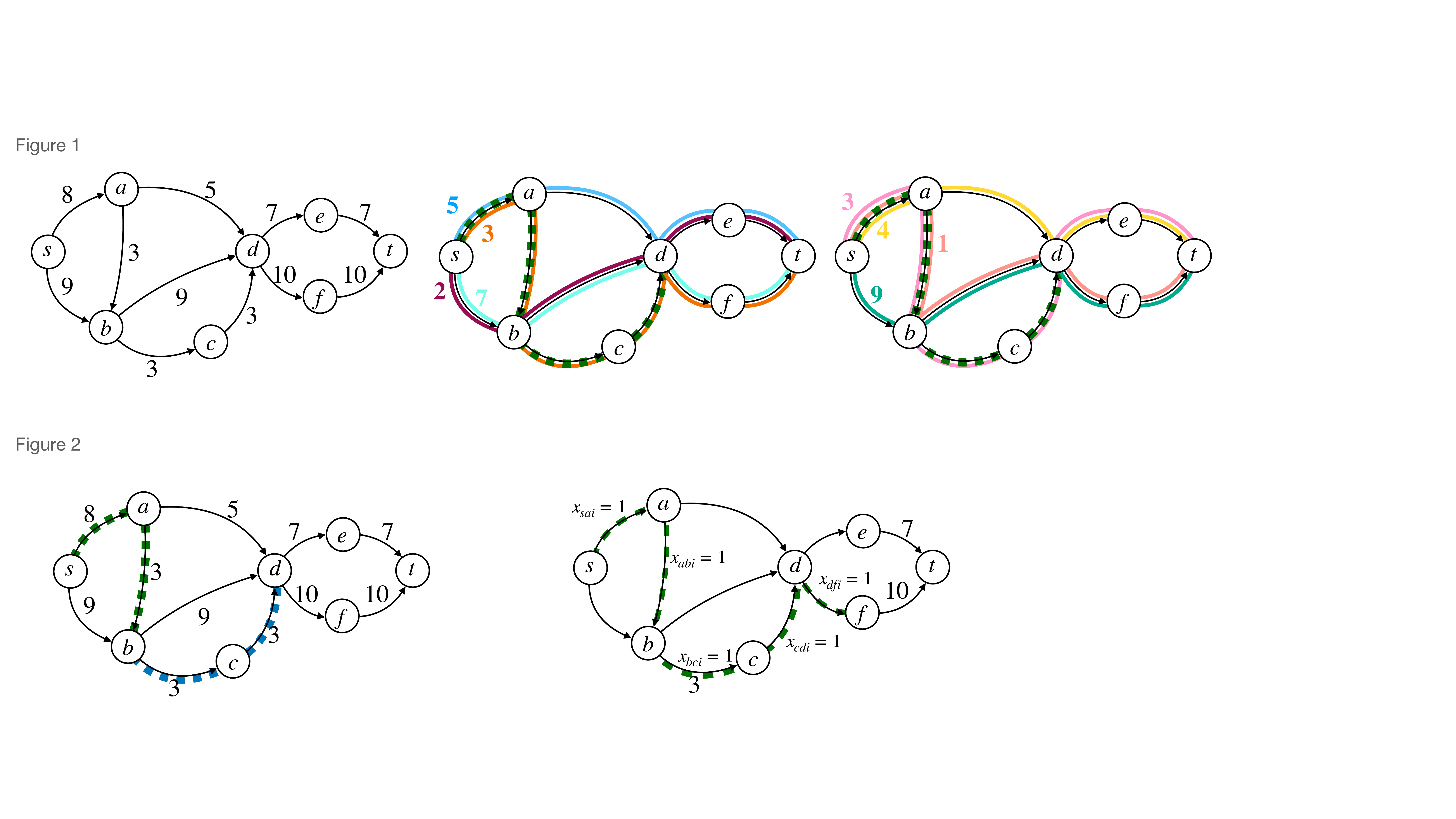}
\caption{The two subpaths (red and blue) of the green dashed path that are maximal safe paths for all FDs.\label{fig:nonsafe}}
\end{subfigure}
\caption{Flow decompositions and safe paths. The flow network in \subref{fig:flow-network-example} admits different MFDs, in~\subref{fig:FD-example-1} and in~\subref{fig:FD-example-2}. The path $(s,a,b,c,d)$ (dashed green) is a maximal safe path for MFDs, i.e., it is a subpath of some path of all MFDs and it cannot be extended without losing this property. However, the path $(s,a,b,c,d)$ is not safe for all FDs. Indeed, its two subpaths $(s,a,b)$ (dashed red in \subref{fig:nonsafe}) and $(b,c,d)$ (dashed blue in \subref{fig:nonsafe}) are maximal safe paths for all FDs. To see this, note that the excess flow of $(s,a,b)$ is 3, while the excess flow of $(s,a,b,c)$ (and of $(s,a,b,c,d)$) is $-6$.\label{fig:safeMFD}}
\end{figure}

\paragraph{\textbf{Flow decomposition.}} In the flow decomposition problem we are given a flow network $(V,E,f)$, where $G = (V,E)$ is a (directed) graph with unique source $s \in V$ and unique sink $t \in V$, and $f$ assigns a positive integer \emph{flow value} $f_{uv}$ to every edge $(u,v) \in E$. \emph{Flow conservation} must hold for every node different from $s$ and $t$, namely, the sum of the flow values entering the node must equal the sum of the flow values exiting the node. See \Cref{fig:flow-network-example} for an example. We say that $k$ $s$-$t$ paths $P_1,\dots,P_k$, with associated positive integer weights $w_1,\dots,w_k$, are a \emph{flow decomposition (FD)} if their superposition equals the flow $f$. Formally, for every $(u,v) \in E$ it must hold that
\begin{equation}
\label{eqn:flow_eq}
\sum_{\substack{i \in \{1,\dots,k\} \text{ s.t. } \\ (u,v) \in P_i}} \hspace{-0.5cm}w_i = f_{uv}.
\end{equation}
See \Cref{fig:FD-example-1,fig:FD-example-2} for two examples. The number $k$ of paths is also called the \emph{size} of the flow decomposition. In the \emph{minimum flow decomposition (MFD) problem}, we need to find a flow decomposition of minimum size.\footnote{In this paper we work only with integer flow values and weights for simplicity and since this is the most studied version of the problem, see e.g.,~\cite{kloster2018practical}. However, the problem can also be defined with fractional weights~\cite{pertea2015stringtie}, and in this case the two problems can have different minima on the same input~\cite{VATINLEN20081390}. This fractional case can also be modeled by ILP~\cite{dias2022fast}, and all the results from our paper also immediately carry over to this variant.} On DAGs, a flow decomposition into paths always exists \cite{ahuja1988network}, but in general graphs, cycles may be necessary to decompose the flow (see e.g.~\cite{dias2022minimum} for different possible formulations of the problem).

For concreteness, we now describe the ILP models from~\cite{dias2022fast} for finding a flow decomposition into $k$ weighted paths in a DAG. They consist of (i) modeling the $k$ paths via the $x_{uvi}$ variables (with constraints \eqref{eqn:flow_conservation}), (ii) adding path-weight variables $w_1,\dots,w_k$, and (iii) requiring that these weighted paths form a flow decomposition, via the following (non-linear) constraint:
\begin{align}
\label{eqn:flow_superposition}
& \sum_{i \in \{1,\dots,k\}} x_{uvi}w_i = f_{uv}, && \forall (u,v) \in E.
\end{align}
This constraint can then be easily linearized by introducing additional variables and constraints; see e.g.~\cite{dias2022fast} for these technical details. However, as mentioned above, the precise formulation of the ILP model $\model$ for a problem is immaterial for our method. Only the two assumptions on $\model$ made above matter for obtaining our results.

\paragraph{\textbf{Safety.}} Given a problem on a graph $G$ whose solutions consist of $k$ paths in $G$, we say that a path $P$ is \emph{safe} if for any solution $P_1,\dots,P_k$ to the problem, there exists some $i \in \{1,\dots,k\}$ such that $P$ is a subpath of $P_i$. If the problem is given as an ILP model $\model$, we also say that $P$ is \emph{safe for $\model$}. We say that $P$ is a \emph{maximal safe path}, if $P$ is a safe path and there is no larger safe path containing $P$ as subpath.
\cite{khan2022safety} characterized safe paths for \emph{all} FDs (not necessarily of minimum size) using the \emph{excess flow} $f_P$ of a path $P$, defined as the flow on the first edge of $P$ minus the flow on the edges out-going from the internal nodes of $P$, and different from the edges of $P$ (see \Cref{fig:nonsafe} for an example). It holds that $P$ is safe for all FDs if and only if $f_P > 0$~\cite{khan2022safety}. The excess flow can be computed in time linear in the length of $P$ (assuming we have pre-computed the flow outgoing from every node), giving thus a linear-time verification of whether $P$ is safe.

A basic property of safe solutions is that any sub-solution of them is also safe. Computing safe paths for MFDs can thus potentially lead to joining several safe paths for FDs, obtaining longer paths from the unknown sequences we are trying to assemble. See \Cref{fig:safeMFD} for an example of a maximal safe path for MFDs and two maximal subpaths of it that are safe for FDs.

\subsection{Finding Maximal Safe Paths for MFD via ILP}

We now present a method for finding all maximal safe paths for MFD via ILP. The basic idea is to define an inner ``safety test'' which can be repeatedly called as part of an outer algorithm over the entire instance to find all maximal safe paths. Because calls to the ILP solver are expensive, the guiding choice for our overall approach is to minimize the number of ILP calls. This inspires us to test the safety of a \emph{group} of paths as the inner safety test, which we achieve by augmenting our ILP model so that it can give us information about the safety of the paths in the set.  We use this to define a recursive algorithm to fully determine the safety status of each path in a group of paths. We can then structure the safety test in either a top-down manner (starting with long unsafe paths and shrinking them until they are safe) or a bottom-up manner (starting with short safe paths and lengthening them until they become unsafe).

\subsubsection{Safety test (inner algorithm)}
\label{sec:safety_test}

Let $\model(\vars,\cons)$ be an ILP model as discussed in \Cref{sec:preliminaries}; namely, its $k$ solution paths are modeled by binary variables $x_{uvi}$ for each $(u,v) \in E$ and each $i \in \{1,\dots,k\}$. We assume that $\model(\vars,\cons)$ is feasible (i.e., the problem admits at least one solution). We first show how to modify the ILP model so that, for a given set of paths, it can tell us one of the following: (1) a set of paths that are \emph{not safe} (the remaining being of unknown status), or (2) that all paths are safe.
The idea is to maximize the number of paths that can be simultaneously avoided from the given set of paths.

Let $\mathcal{P}$ be a set of paths. For each path $P \in \mathcal{P}$, we create an auxiliary binary variable $\gamma_P$ that indicates:
\begin{equation}\label{eq:group_safety}
\gamma_P \equiv \begin{cases}
1 & \text{if $P$ was avoided in the solution},\\
0 & \text{otherwise}.
\end{cases}
\end{equation}

Since the model solutions are \emph{paths} (i.e., not repeating nodes), we can encode whether $P$ appears in the solution by whether \emph{all} of the $\ell - 1$ edges of $P$ appear simultaneously.
Using this fact, we add a new set of constraints $\rcon(\mathcal{P})$ that include the $\gamma_P$ indicator variables for each path $P \in \mathcal{P}$:
\begin{align}
\label{eq:group-testing-constraints}
& \rcon(\mathcal{P}) := \{x_{v_1v_2i} + x_{v_2v_3i} + \cdots + x_{ v_{\ell-1} v_\ell i}  \nonumber \\
& \quad \leq \ell - 1 - \gamma_P: \forall i \in \{1,\dots,k\},\forall P \in \mathcal{P}\}.
\end{align}

Next, as the objective function of the ILP model, we require that it should maximize the number of avoided paths from $\mathcal{P}$, i,e., the sum of the $\gamma_P$ variables:
\begin{equation}
    \max \sum_{P \in \mathcal{P}} \gamma_P.
    \label{eqn:maxavoid}
\end{equation}
All paths $P$ such that $\gamma_P = 1$ are \emph{unsafe}, since they were avoided in some
minimum flow decomposition.  Conversely, if the objective value of \cref{eqn:maxavoid} was $0$, then $\gamma_P = 0$ for all paths in $\mathcal{P}$, and it must be that all paths in $\mathcal{P}$ are safe (if not, at least one path
could be avoided and increase the objective).
We encapsulate this group testing ILP in a function
\textsf{GroupTest}$(\model,\mathcal{P})$ that returns a set $\mathcal{N} \subseteq \mathcal{P}$
with the properties that: (1)  if $\mathcal{N} = \emptyset$, then all paths in $\mathcal{P}$ are safe, and (2) if $\mathcal{N} \neq \emptyset$, then all paths in $\mathcal{N}$ are unsafe (and $|\mathcal{N}|$ is maximized).

\begin{figure}
    \allowdisplaybreaks
    \centering
    \includegraphics[scale=0.25]{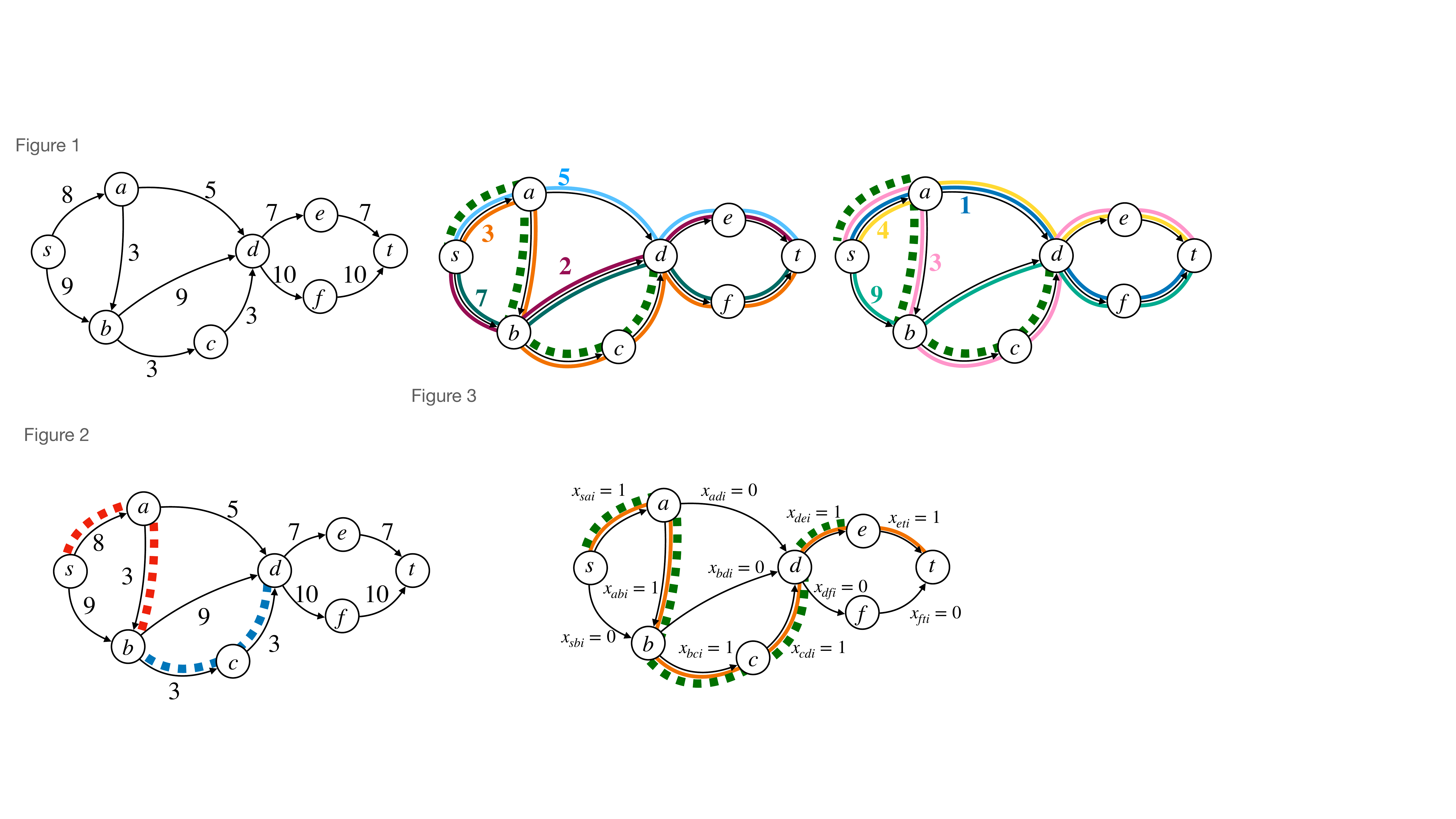}
    \caption{
    Illustration of modeling a solution path and a tested path via binary edge variables and safety verification constraints. The $i$\textsuperscript{th}
    solution path $P_i$ is shown in orange, and a tested path $P$ is shown in dashed green.
    Constraint \eqref{eq:group-testing-constraints} includes $x_{sai} + x_{abi} + x_{bci} + x_{cdi} + x_{dei} \leq 5 - \gamma_P$.  This simplifies to $\gamma_P \leq 0$, thus forcing $\gamma_P = 0$, which indicates $P$ was not avoided in the solution.
    \label{fig:safety-binary}
    }
\end{figure}

We employ \textsf{GroupTest}$(\model,\mathcal{P})$ to construct a recursive procedure \textsf{GetSafe}$(\model,\mathcal{P})$ that
determines all safe paths in $\mathcal{P}$, as shown in \Cref{algo:pathset-safety}.

\begin{algorithm}[t]
\allowdisplaybreaks
\SetKw{kwReturn}{return}
\KwIn{A feasible ILP model $\model(\vars,\cons)$, and a set of paths $\mathcal{P}$}
\KwOut{Those paths $P \in \mathcal{P}$ that are safe for $\model(\vars,\cons)$}

\SetKwProg{myproc}{Procedure}{}{}
\myproc{\textsf{GetSafe}$(\model,\mathcal{P})$}{
$\mathcal{N}$ = \textsf{GroupTest}$(\model,\mathcal{P})$\
\eIf{$\mathcal{N} = \emptyset$}{
    \kwReturn $\mathcal{P}$\
}
{
    \kwReturn \textsf{GetSafe}$(\model,\mathcal{P} \setminus \mathcal{N})$\
}
}
\caption{Testing a set of paths $\mathcal{P}$ for safety\label{algo:pathset-safety}.}
\end{algorithm}

We note that in the special case that $|\mathcal{P}|=1$, \textsf{GetSafe}$(\model,\mathcal{P})$ makes only a single call to
the ILP via \textsf{GroupTest}$(\model,\mathcal{P})$ to determine whether not the given path is safe. With this safety test for a single path, we can easily adapt a standard two-pointer approach as the outer algorithm to find all maximal safe paths for MFD by starting with some MFD solution $P_1,\dots,P_k$ of $\model(\vars,\cons)$. This same procedure was used in \cite{Khan:2022wo} to find all maximal safe paths for FD, using an excess flow check as the inner safety algorithm.

\subsubsection{Find all maximal safe paths (outer algorithm)}

We give two algorithms for finding all maximal safe paths.  Both algorithms use a similar approach, however the first uses a top-down approach starting from the original full solution paths and reports all safe paths (these again must be maximal safe), and then trims all the unsafe paths to find new maximal safe paths. The second is bottom-up in that it tries to extend known safe subpaths until they cannot be further extended (and at this point must be maximal safe).  We present the first algorithm in detail and defer discussion of the second to the appendix.

We say a set of subpaths $\mathcal{T} = \{ P_i[l_i, r_i] \}$ is a \emph{trimming core} provided that for any unreported maximal safe path $P = P_i[l,r]$, there is a $P_i[l_i, r_i] \in \mathcal{T}$,
where $l_i \le l \le r \le r_i$.

\begin{algorithm}[]
\allowdisplaybreaks
\SetKw{kwContinue}{continue}
\SetKw{kwReturn}{return}
\SetKw{kwOutput}{output}
\SetKw{kwAnd}{and}
\SetKw{kwOr}{or}
\SetKw{kwNot}{not}
\SetKw{kwInfeasible}{infeasible}
\SetKw{kwFeasible}{feasible}
\SetKwRepeat{Do}{repeat}{until}
\KwIn{An ILP model $\model$ and a trimming core set $\mathcal{T}$}
\KwOut{All maximal safe paths for $\model$ that are trimmed subpaths of $\mathcal{T}$}

\SetKwProg{myproc}{Procedure}{}{}
\myproc{\textsf{AllMaxSafe-TopDown}$(\model,\mathcal{T}$)}{
$\mathcal{S}$ = \textsf{GetSafe}$(\model,\mathcal{T})$\
\For{$P_i[l_i,r_i] \in \mathcal{S}$}{
        \kwOutput $P_i[l_i,r_i]$\
}
$\mathcal{U} = \mathcal{T} \setminus \mathcal{S}$\
$\mathcal{L} = \{ P_i[l_i+1,r_i] : P_i[l_i,r_i] \in \mathcal{U}, (r_i = |P_i|\ \kwOr \ P_i[l_i+1, r_i+1] \in \mathcal{U}) \}$\
$\mathcal{R} = \{ P_i[l_i,r_i-1] : P_i[l_i,r_i] \in \mathcal{U}, (l_i = 1\ \kwOr \ P_i[l_i-1, r_i-1] \in \mathcal{U})\}$\
$\mathcal{P} = \mathcal{L} \cup \mathcal{R}$\
\If{$\mathcal{P} \ne \emptyset$}
    {
        \textsf{AllMaxSafe-TopDown}$(\model,\mathcal{P})$\
    }
}

\caption{An algorithm to compute all maximal safe paths that can be trimmed from a trimming core set $\mathcal{T}$.\label{algo:allmax-topdown}}
\end{algorithm}

We will use the original $k$ solution paths $\{P_i\}$ as our initial trimming core; the complete algorithm is given in \cref{algo:allmax-topdown}. See \cref{fig:grouptesting} in the appendix for an illustration of the algorithm's initial steps. The algorithm first checks if any of the paths in $\mathcal{T}$ are safe; if so, these are reported as maximal safe.  For those paths that were unsafe, it then considers trimming one vertex from the left and one vertex from the right to create new subpaths.  Of these subpaths, some may  be contained in a safe path in $\mathcal{T}$; these subpaths can be ignored as they are not maximal safe.  The algorithm recurses on those subpaths whose safety status cannot be determined (lines $6$--$10$).  In this way, the algorithm maintains the invariant that no
paths in $\mathcal{T}$ are properly contained in a safe path; thus paths reported in line $4$ must be maximal safe.

\section{Experiments}
To test the performance of our methods, we computed safe paths using different safety approaches and reported the quality and running time performances as described below. Additional details on the experimental setup are given in the appendix.

\paragraph{\textbf{Implementation details -- SafeMFD.}}
We implemented the previously described algorithms to compute all maximal safe paths for minimum flow decompositions in \texttt{Python}. The implementation, \emph{SafeMFD}, uses the package \texttt{NetworkX}~\cite{hagberg2008exploring} for graph processing and the package \texttt{gurobipy}~\cite{gurobi} to model and solve the ILPs and it is openly available\footnote{\url{https://github.com/algbio/mfd-safety}}. Our fastest variant (see \cref{tab:performance_variants} in the appendix for a comparison of running times) implements \Cref{algo:allmax-topdown} using the group testing in \Cref{algo:pathset-safety}. We used this variant to compare against other safety approaches. All tested variants of \emph{SafeMFD} implement the following two optimizations:
\begin{enumerate}
    \item~Before processing an input flow graph we contract it using Y-to-V contraction~\cite{tomescu2017safe}, which is known~\cite{kloster2018practical} to maintain (M)FD solution paths. Moreover, since edges in the contracted graph correspond to extended unitigs~\cite{MGMB07,jacksonthesis,kingsford2010assembly}, source-to-sink edges are further removed from the contracted graph and reported as safe. As such, our algorithms compute all maximal safe paths for funnels~\cite{garlet2020efficient,Khan:2022wo} without using the ILP.
    \item~Before testing the safety of a path we check if its \emph{excess-flow}~\cite{Khan:2022wo} is positive. If this is the case, the path is removed from the corresponding test. Having positive excess flow implies safety for \emph{all} flow decomposition and thus also safety for minimum flow decompositions.
\end{enumerate}

\paragraph{\textbf{Safety approaches tested.}}
We compare the following state-of-the-art safety approaches:
\begin{description}
\item[\emph{EUnitigs:}~]
Maximal paths made up of a prefix of nodes with in-degree one followed by nodes with out-degree one; also called \emph{extended unitigs}~\cite{tomescu2017safe,MGMB07,jacksonthesis,kingsford2010assembly}. We use the C++ implementation provided by Khan et~al.~\cite{Khan:2022wo} (which computes only the extended unitigs contained in FD paths).
\item[\emph{SafeFlow:}~]
Maximal safe paths for all flow decompositions~\cite{Khan:2022wo}. We use the C++ implementation provided by Khan et~al.~\cite{Khan:2022wo}.
\item[\emph{SafeMFD:}~]
Maximal safe paths for all minimum flow decompositions, as proposed in this work.
Every flow graph processed is given a \emph{time budget} of 2 minutes. If a flow graph consumes its time budget, the solution of \emph{SafeFlow} is output instead.
\item[\emph{SafeEPC:}~]
Maximal safe paths for all constrained path covers of edges. Previous authors \cite{Caceres:2022ve,Khan:2022wo} have considered safe path covers of the nodes, but for a more fair comparison, we instead use path covers of edges. To this end, we transform the input graphs by splitting every edge by adding a node in the middle and run the C++ implementation provided by the authors of~\cite{Caceres:2022ve}. Since flow decompositions are path covers of edges, safe paths for \emph{all} edge path covers are subpaths of safe paths for MFD. However, we restrict the path covers to those of minimum size and minimum size plus one, as recommended by the authors of~\cite{Caceres:2022ve} to obtain good coverage results while maintaining high precision.
\end{description}

All safety approaches require a post processing step for removing duplicates, prefixes and suffixes. We use the C++ implementation provided by \cite{Khan:2022wo} for this purpose.

\paragraph{\textbf{Datasets.}}
We use two datasets of flow graphs inspired by RNA transcript assembly. The datasets were created by simulating abundances on a set of transcripts and then \emph{perfectly} superposing them into a splice graphs that are guaranteed to respect flow conservation. As such, the ground truth corresponds to a flow decomposition (not necessarily minimum). To avoid a skewed picture of our results we filtered out trivial instances with a unique flow decomposition (or funnels, see~\cite{garlet2020efficient,Khan:2022wo}) from the two datasets.\footnote{The exact datasets used in our experiments can be found at \url{https://zenodo.org/record/7182096}.}
\begin{description}
\item[Catfish:~] Created by \cite{shao2017theory}, it includes 100 simulated human, mouse and zebrafish transcriptomes using Flux-Simulator~\cite{griebel2012modelling} as well as 1,000 experiments from the Sequence Read Archive simulating abundances using Salmon~\cite{patro2015salmon}. We took one experiment per dataset, which corresponds to 27,696 non-trivial flow graphs.
\item[RefSim:~] Created by \cite{Caceres:2022ve} from the Ensembl~\cite{yates2020ensembl} annotated transcripts of GRCh.104 \emph{homo sapiens} reference genome, and later augmented by Khan et~al.~\cite{Khan:2022wo} with simulated abundances using the RNASeqReadSimulator~\cite{li2014rnaseqreadsimulator}. This dataset has 10,323 non-trivial graphs.
\end{description}

\paragraph{\textbf{Quality metrics.}}

We use the same quality metrics employed by previous multi-assembly safety approaches~\cite{Caceres:2022ve,Khan:2022wo}. We provide a high-level description of them for completeness.

\begin{description}
\item[Weighted precision of reported paths:~] As opposed to normal precision, the weighted version considers the length of the reported subpaths. It is computed as the total length of the correctly reported subpaths divided by the total length of all reported subpaths. A reported subpath is considered \emph{correct} if and only if it is a subpath of some path in the ground truth (exact alignment of exons/nodes).
\item[Maximum coverage of a ground truth path $P$:~] The longest segment of $P$ covered by some reported subpath (exact alignment of exons/nodes), divided by $|P|$.
\end{description}

We compute the \textbf{weighted precision of a graph} as the average weighted precision over all reported paths in the graph, and the \textbf{maximum coverage of a graph} as the average maximum coverage over all ground truth paths in the graph.

\begin{description}
\item[F-Score of a graph:~] Harmonic mean between weighted precision and maximum coverage of a graph, which assigns a global score to the corresponding approach on the graph.
\end{description}

These metrics are computed per flow graph and reported as an average. In the case of the \emph{Catfish} dataset the metrics are computed in terms of exons (nodes), since genomic coordinates of exons are missing, whereas in the case of the \emph{RefSim} dataset the metrics are computed in terms of genomic positions, as this information is present in the input.

\section{Results and Discussion}

\begin{table}
\caption{Summary of quality metrics for both datasets. For \emph{Catfish}, the metrics are computed in terms of nodes/exons and for \emph{RefSim} in terms of genomic positions; $t$ is the number of ground truth paths.\label{tab:both_datasets_quality}}
\centering
\begin{tabular}{|c|c|c|c|c|c|c|c|}
\hline
Dataset & Graphs & Algorithm & Max. Coverage & Wt. Precision &  F-Score\\ \hline
\multirow{4}{*}{\parbox[c]{2cm}{\begin{center}\emph{Catfish}\end{center}}}
& \multirow{4}{*}{\parbox[c]{2cm}{\begin{center}All\\(100\%)\end{center}}}
& EUnitigs & 0.60 & 1.00 & 0.74\\
& & SafeEPC & 0.60 & 0.99 & 0.74\\
& & SafeFlow & 0.71 & 1.00 & 0.82\\
& & SafeMFD & 0.88 & 0.99 & 0.93\\
\hline
\multirow{12}{*}{\parbox[c]{2cm}{\begin{center}\emph{RefSim}\end{center}}}
& \multirow{4}{*}{\parbox[c]{2cm}{\begin{center}$t\le 10$\\(68\%)\end{center}}}
& EUnitigs & 0.72 & 1.00 & 0.83\\
& & SafeEPC & 0.73 & 1.00 & 0.84\\
& & SafeFlow & 0.84 & 1.00 & 0.91\\
& & SafeMFD & 0.97 & 0.99 & 0.98\\
\cline{2-6}
& \multirow{4}{*}{\parbox[c]{2cm}{\begin{center}$t \le 15$\\(84\%)\end{center}}}
& EUnitigs & 0.70 & 1.00 & 0.82\\
& & SafeEPC & 0.71 & 1.00 & 0.83\\
& & SafeFlow & 0.83 & 1.00 & 0.90\\
& & SafeMFD & 0.96 & 0.98 & 0.97\\
\cline{2-6}
& \multirow{4}{*}{\parbox[c]{2cm}{\begin{center}All\\(100\%)\end{center}}}
& EUnitigs & 0.68 & 1.00 & 0.80\\
& & SafeEPC & 0.69 & 0.99 & 0.81\\
& & SafeFlow & 0.81 & 1.00 & 0.89\\
& & SafeMFD & 0.93 & 0.91 & 0.90\\
\hline
\end{tabular}
\end{table}

In the Catfish dataset, \emph{EUnitigs} and \emph{SafeFlow} ran in less than a second, while \emph{SafeEPC} took approximately 30 seconds to compute. On the other hand, solving a harder problem, \emph{SafeMFD} took approximately 1.5 hours to compute in the rest of the dataset, timing out in only 54 graphs (we use a cutoff of 2 minutes), i.e., only 0.2\% of the entire dataset. This equates to only 0.2 seconds on average per solved graph, underlying the scalability of our approach.

\Cref{tab:both_datasets_quality} shows that \emph{SafeMFD}, on average, covers close to 90\% of the ground truth paths, while maintaining a high precision (99\%). This corresponds to an increase of approximately 25\% in coverage against its closest competitor \emph{SafeFlow}. \emph{SafeMFD} also dominates in the combined metric of F-Score, being the only safe approach with F-Score over 90\%. \Cref{fig:catfish_quality_nodes} in the appendix shows the metrics on graphs grouped by number $t$ of ground truth paths, indicating the dominance in coverage and F-Score of \emph{SafeMFD} across all values of $t$, and indicating that the decrease in precision appears for large values of $t$ ($t\ge 12$).

In the harder RefSim dataset, \emph{EUnitigs} and \emph{SafeFlow} also ran in less than a second, while \emph{SafeEPC} took approximately 2 minutes. In this case, \emph{SafeMFD} ran out of time in 1,562 graphs (15\% of the entire dataset); however, recall that in these experiments we allow a time budget of only 2 minutes. In the rest of the dataset, it took approximately 7.5 hours in total, corresponding to only 3 seconds on average per graph, again underlying that our method, even though it solves many NP-hard problems \emph{for each input graph}, overall scales sufficiently well.

\Cref{tab:both_datasets_quality} shows that again \emph{SafeMFD} dominates in coverage, being the only approach obtaining coverage over 90\%, with is a 15\% improvement over \emph{SafeFlow}. This time its precision drops to close to 90\%, and obtaining an F-Score of 90\%, very similar to its closest competitor, \emph{SafeFlow}. However, recall that coverage is computed only from \emph{correctly aligned paths}, thus the drop in precision comes only from safe paths not counting in the coverage metric. If we restrict the metrics to graphs with at most 15 ground truth paths, which is still a significant proportion (84\%) of the entire dataset, then \emph{SafeMFD} has a very high precision (98\%) while improving coverage by 15\% with respect to \emph{SafeFlow}. Thus, the drop in precision occurs in graphs with a large number of ground truth paths, which can also be corroborated by \Cref{fig:refsim_quality_positions} in the appendix.

These drops in precision (both in RefSim and Catfish) for large $t$ can be explained by the fact that a larger number of ground truth paths produces more complex splice graphs and introduces more artificial solutions of potentially smaller size. As such, the larger $t$, the less likely that the ground truth is a minimum flow decomposition of the graph, and thus the more likely that \emph{SafeMFD} reports incorrect solutions. This motivates future work on safety not only on \emph{minimum} flow decompositions but  also in flow decompositions of at most a certain size, analogously to how it is done for \emph{SafeEPC}. This is still easily achievable with our framework by just changing the ILP blackbox, and keeping everything else unchanged (e.g., the inner and outer algorithms). Namely, instead of formulating the ILP model $\model(\vars,\cons)$ to admit solutions of exactly optimal $k$ paths, it can be changed to allow solutions of \emph{at most} some $k'$ paths, with $k'$ greater than the optimal $k$. If $k'$ is also greater than the number of ground truth paths in these complex graphs, then safe paths are fully correct, meaning that we overall increase precision.

\section{Conclusion}

RNA assembly is a difficult problem in practice, with even the top tools reporting low precision values. While there are still many issues that can introduce uncertainty in practice, we can now provide a major source of additional information during the process: which RNA fragments must be included in \emph{any} parsimonious explanation of the data? Though others have considered RNA assembly in the safety framework \cite{findingranges,Khan:2022wo}, we are the first to show that safety can be practically used even when we look for optimal (i.e., minimum) size solutions. Our experimental results show that safe paths for MFD clearly outperform other safe approaches for the \emph{Catfish} dataset, commonly used in this field. On a significant proportion of the second dataset, safe paths for MFD still significantly outperforms other safe methods.

More generally, this is the first work to show that the safety framework can be practically applied to NP-hard problems, where the inner algorithm is an efficient test of safety of a group of paths, and the outer algorithm guides the applications of this test.
Because our method was very successful on our test data set, there is strong motivation to try the approach to on other NP-hard graph problems whose solutions are sets of paths. For example, we could study other variations on MFD, such as finding flow decompositions minimizing the longest path (NP-hard when flow values are integer \cite{baier2004flows,pienkosz2015integral}). The approach given in this paper can also be directly extended to find decompositions into both cycles and paths \cite{dias2022minimum}, though not trails and walks, because they repeat edges. We could also formulate a safety test for classic NP-hard graph problems like Hamiltonian path.


\subsubsection*{Acknowledgements}

This work was partially funded by the European Research Council (ERC) under the European Union's Horizon 2020 research and innovation programme (grant agreement No.~851093, SAFEBIO), partially by the Academy of Finland (grants No.~322595, 352821, 346968),
and partially by the US National Science Foundation (NSF) (grants No.~1759522, 1920954).


\bibliography{biblio.bib}
\bibliographystyle{plain}

\newpage
\appendix

\section{Additional Figures}

\begin{figure}
\allowdisplaybreaks
\centering
\begin{subfigure}[c]{0.32\textwidth}
\includegraphics[width=\textwidth]{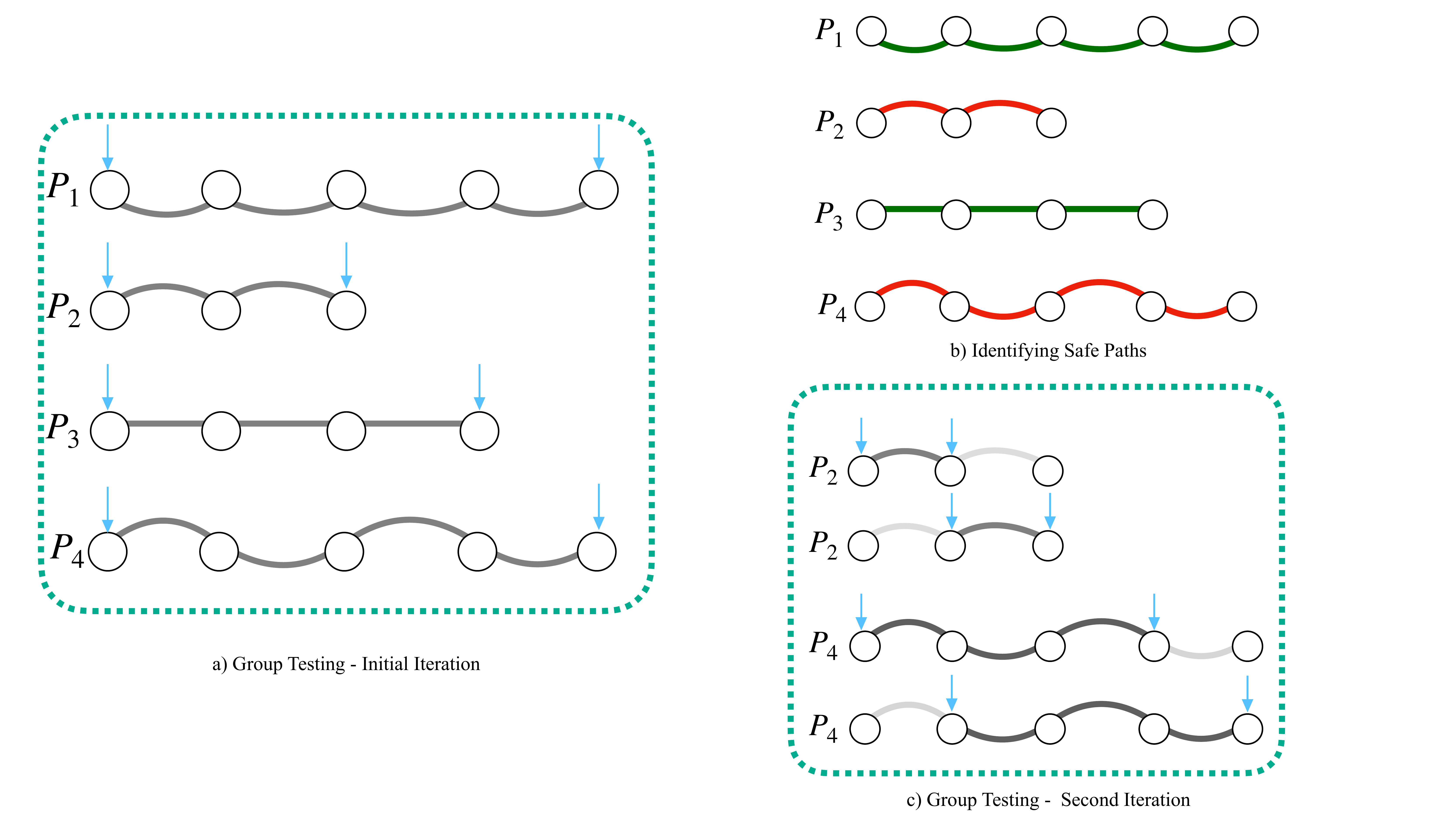}
\caption{First group test\label{fig:gp1}}
\end{subfigure}
\begin{subfigure}[c]{0.33\textwidth}
\includegraphics[width=\textwidth]{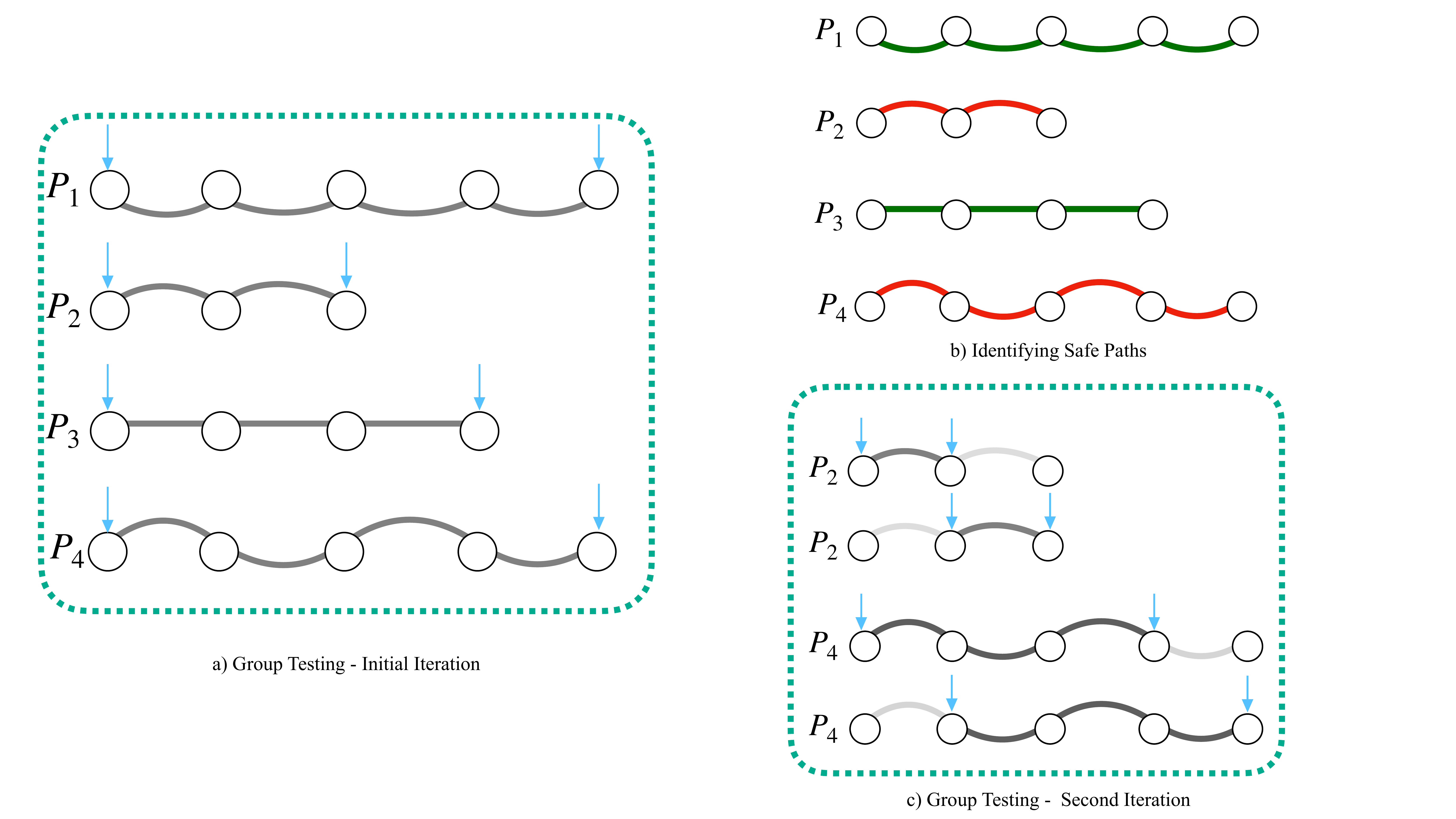}
\caption{Result: $\{P_1, P_3\}$ are safe, $\{P_2, P_4\}$ are unsafe\label{fig:gp2}}
\end{subfigure}
\begin{subfigure}[c]{0.33\textwidth}
\includegraphics[width=\textwidth]{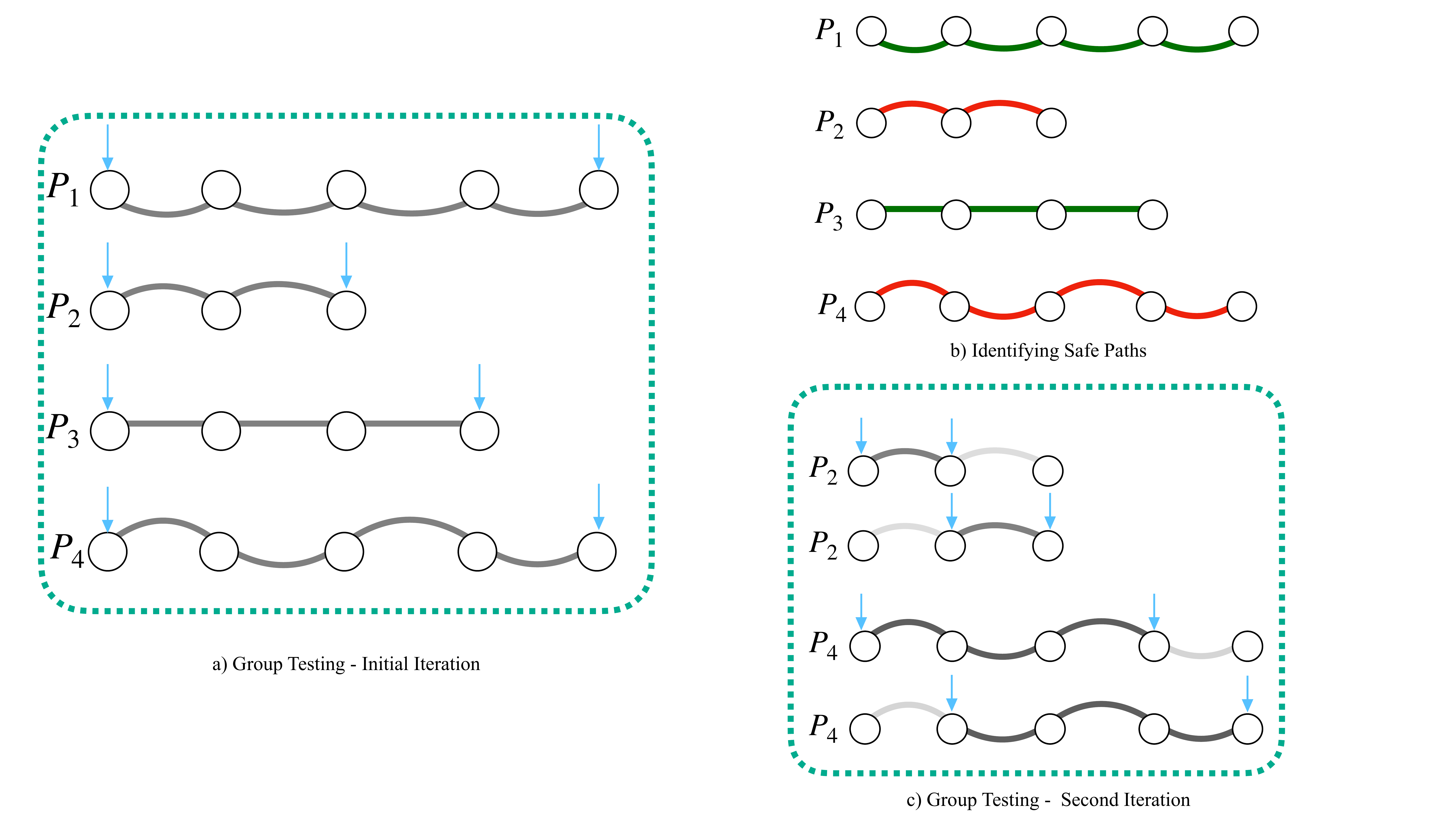}
\caption{Second group test\label{fig:gp3}}
\end{subfigure}
\caption{
Illustration of the initial group tests performed by \cref{algo:allmax-topdown}.  \cref{fig:gp1} shows the first group test (using \cref{algo:pathset-safety}) on MFD solution paths $\{P_1,P_2,P_3,P_4\}$; suppose $\{P_1, P_3\}$ were safe (\cref{fig:gp2}); these are then reported as maximal safe.  In this case we trim $\{P_2, P_4\}$ on both the left and right and make the next group test shown in \cref{fig:gp3}.\label{fig:grouptesting}}
\end{figure}

\begin{figure}[t]
     \centering
     \begin{subfigure}[b]{0.32\textwidth}
         \centering
         \includegraphics[trim=30 25 75 90,clip,  width=\textwidth]{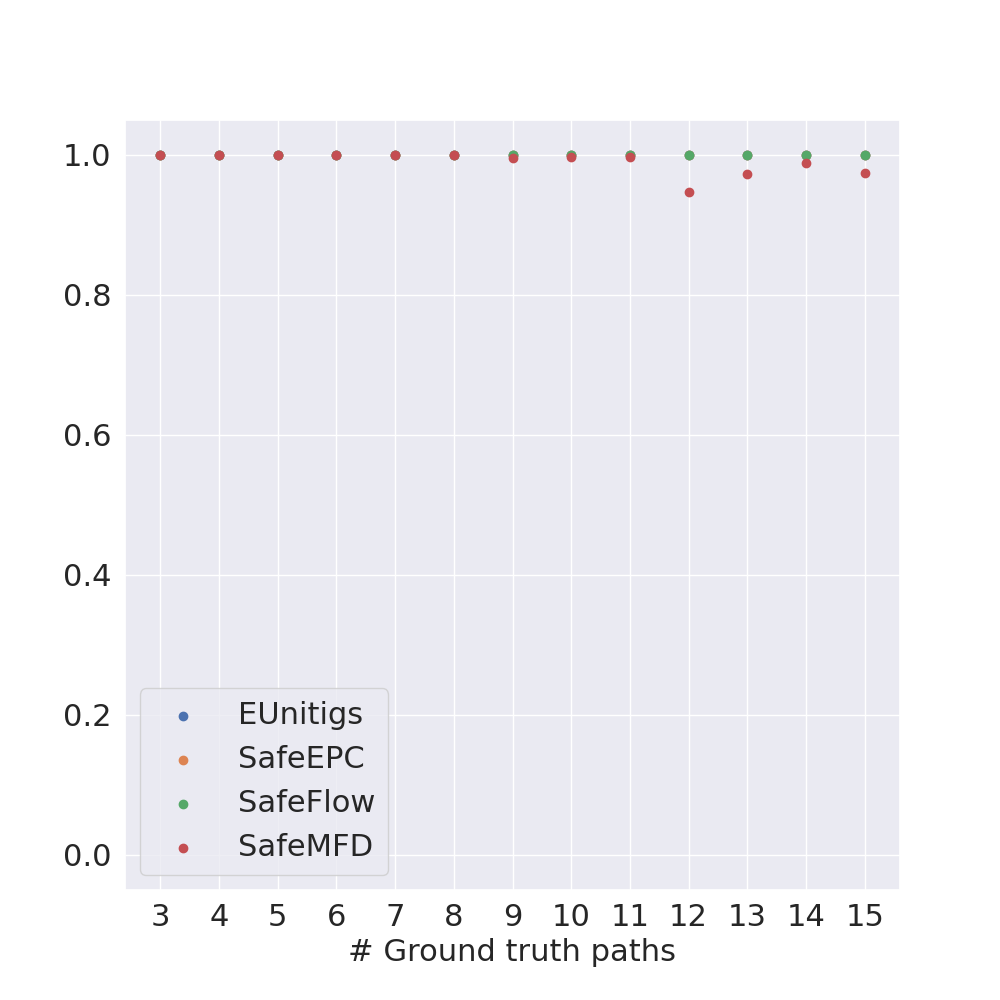}
         \caption{Weighted Precision}
          \label{fig:catfish_weighted_precision_nodes}
     \end{subfigure}
     \begin{subfigure}[b]{0.32\textwidth}
         \centering
         \includegraphics[trim=30 25 75 90,clip,  width=\textwidth]{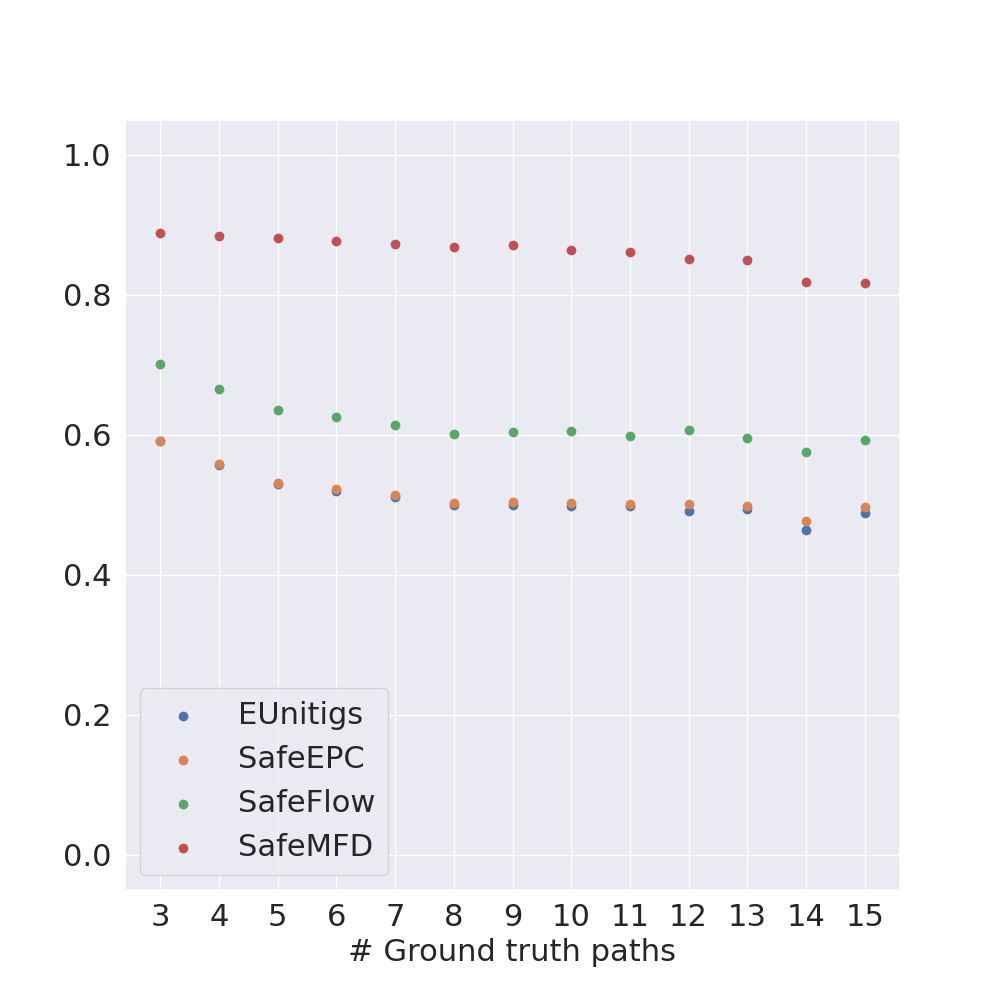}
         \caption{Maximum Coverage}
         \label{fig:catfish_max_coverage_nodes}
     \end{subfigure}
    \begin{subfigure}[b]{0.32\textwidth}
         \centering
         \includegraphics[trim=30 25 75 90,clip,  width=\textwidth]{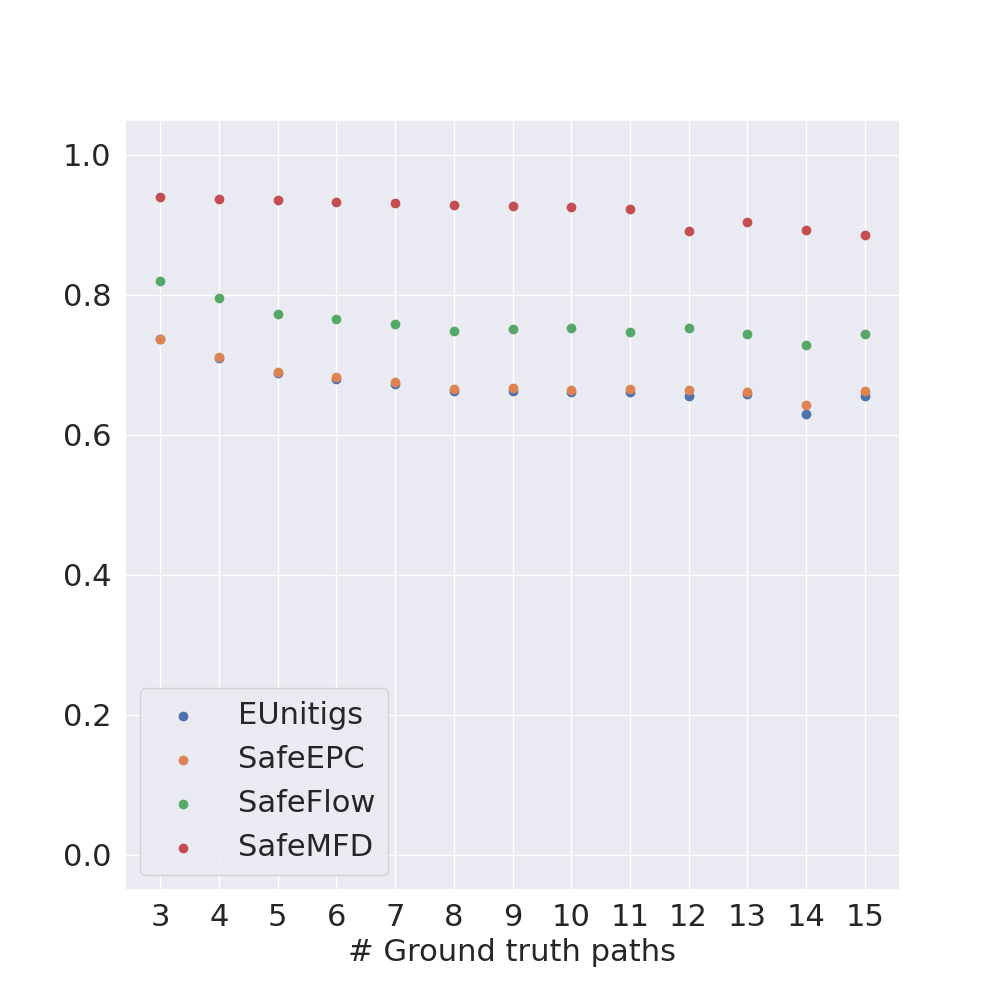}
         \caption{F-Score}
         \label{fig:catfish_f_score_nodes}
     \end{subfigure}
\caption{Quality metrics on graphs distributed by number of paths in the ground truth for the Catfish dataset. The metrics are computed in terms of exons/nodes.}
\label{fig:catfish_quality_nodes}
\end{figure}

\newpage
\begin{figure}[t]
     \centering
     \begin{subfigure}[b]{0.32\textwidth}
         \centering
         \includegraphics[trim=30 25 75 90,clip,  width=\textwidth]{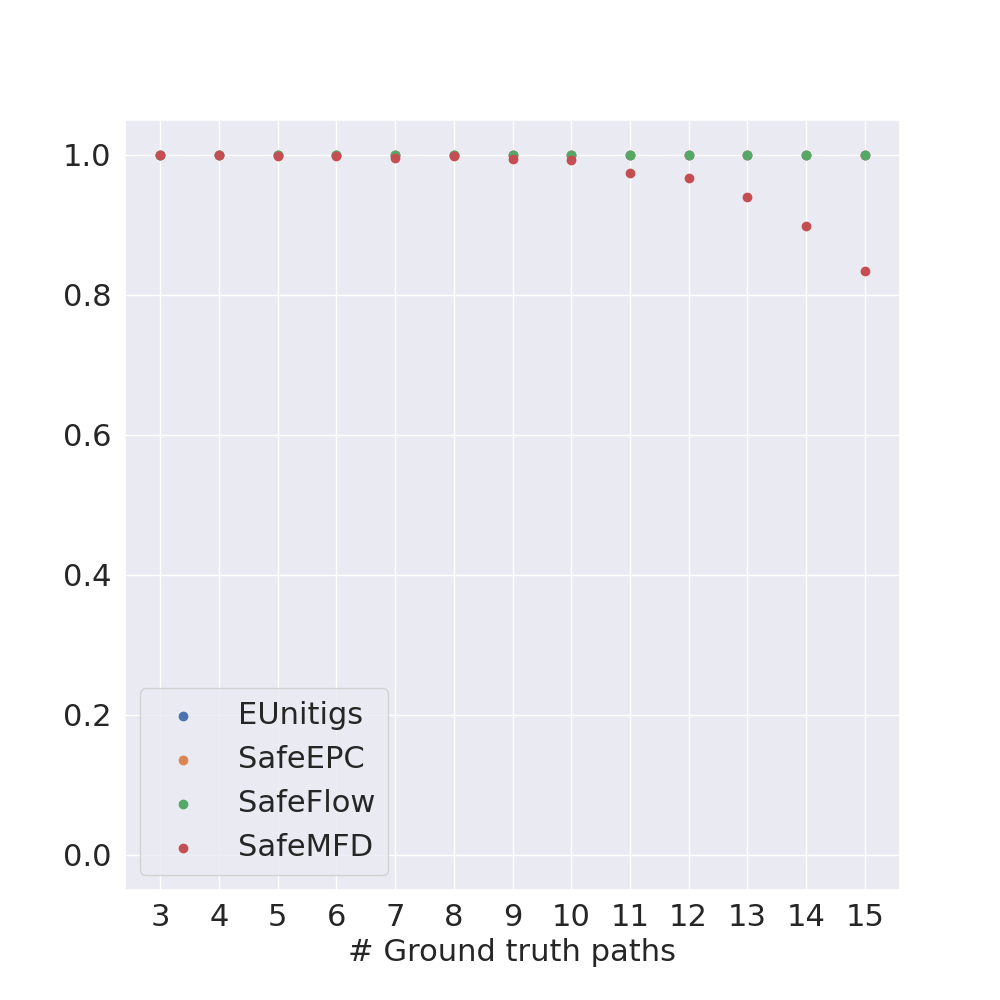}
         \caption{Weighted Precision}
          \label{fig:refsim_weighted_precision_positions}
     \end{subfigure}
     \begin{subfigure}[b]{0.32\textwidth}
         \centering
         \includegraphics[trim=30 25 75 90,clip,  width=\textwidth]{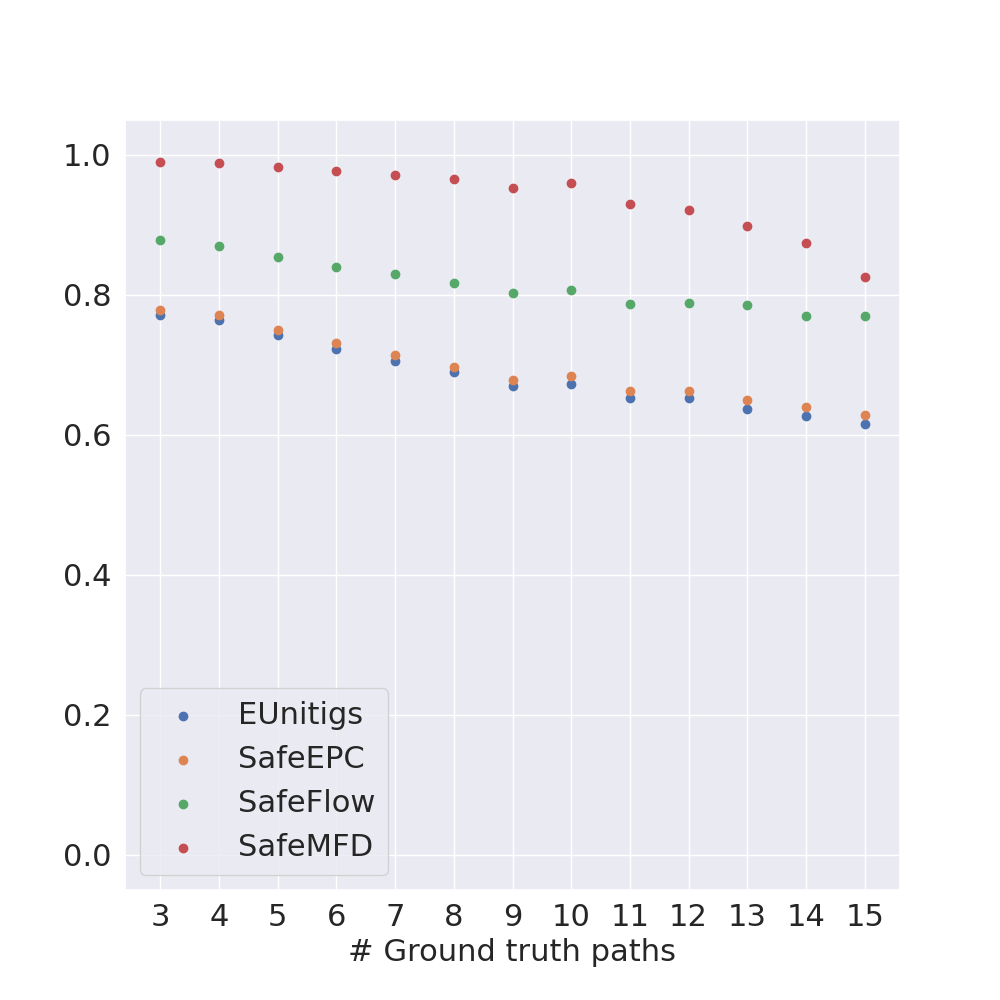}
         \caption{Maximum Coverage}
         \label{fig:refsim_max_coverage_positions}
     \end{subfigure}
    \begin{subfigure}[b]{0.32\textwidth}
         \centering
         \includegraphics[trim=30 25 75 90,clip,  width=\textwidth]{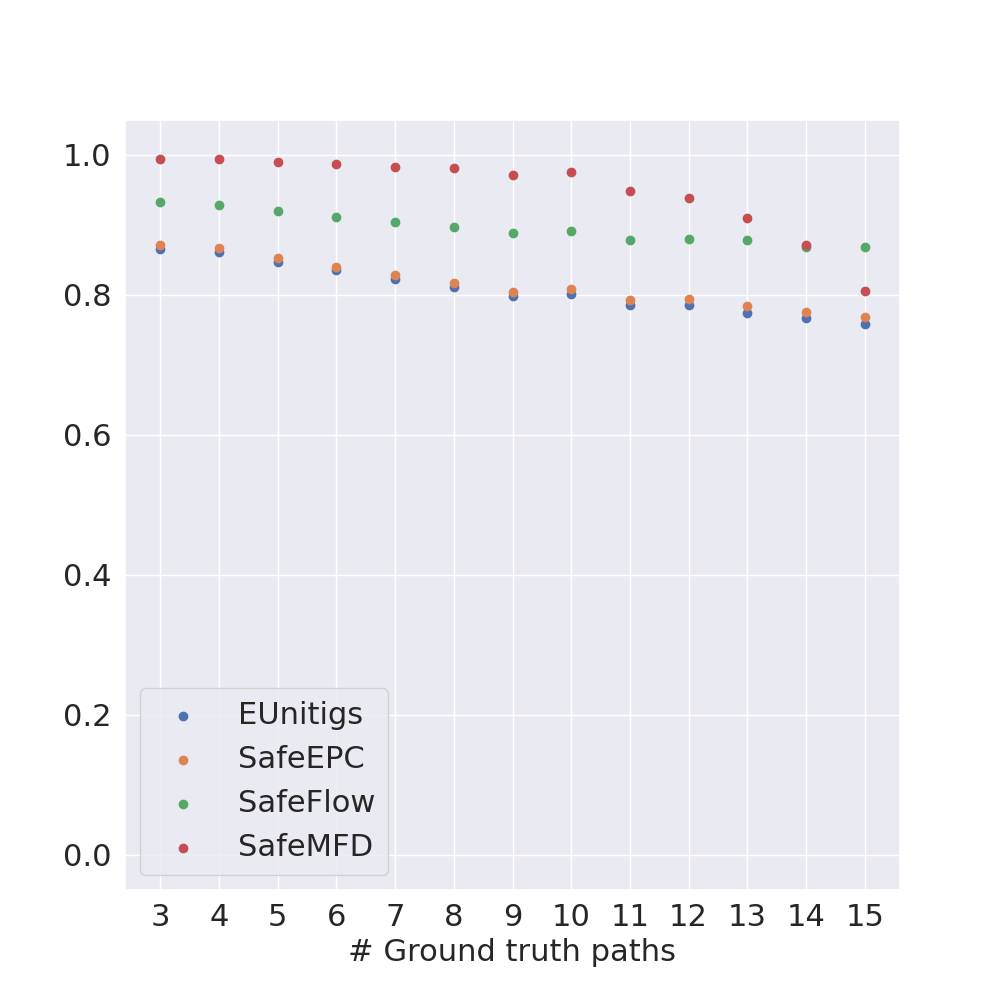}
         \caption{F-Score}
         \label{fig:refsim_f_score_positions}
     \end{subfigure}
\caption{Quality metrics on graphs distributed by number of paths in the ground truth for the RefSim dataset. The metrics are computed in terms of genomic positions.}
        \label{fig:refsim_quality_positions}
\end{figure}

\section{Additional Algorithms and Experimental Results}
\label{sec:additionalalgs}

\subsection{The bottom-up algorithm}
\label{sec:bottomup}

\cref{algo:allmax-bottomup}, detailed below, uses a bottom-up group-testing strategy to find all maximal safe paths.

\begin{definition}
We say a set of subpaths $\mathcal{E} = \{ P_i[l_i, r_i] \}$ is an \emph{extending core} provided all paths in $\mathcal{E}$ are safe and for
any unreported maximal safe path $P= P_i[l,r]$, there is a $P_i[l_i, r_i] \in \mathcal{E}$,
where $l \le l_i \le r_i \le r$.
\end{definition}

Note that maximal FD-safe subpaths provide an extending core
(as well just the set of all single-edge subpaths in each path).  \Cref{algo:allmax-bottomup}
provides an algorithm to find all maximal safe paths based on group testing, starting from an extending core. The idea is to try both left-extending (by one) and right-extending (by one) each subpath in the core; if neither of these extensions are safe, then we know that that core
subpath must be maximal safe.  Testing all extensions can done quickly using \cref{algo:pathset-safety}.  We then recurse on a new core set consisting of those
extensions that were found to be safe.

\begin{algorithm}[th]
\allowdisplaybreaks
\SetKw{kwContinue}{continue}
\SetKw{kwReturn}{return}
\SetKw{kwOutput}{output}
\SetKw{kwAnd}{and}
\SetKw{kwNot}{not}
\SetKw{kwInfeasible}{infeasible}
\SetKw{kwFeasible}{feasible}
\SetKwRepeat{Do}{repeat}{until}
\KwIn{An ILP model $\model$ and an extending core set $\mathcal{E}$}
\KwOut{All maximal safe paths for $\model$ that extend some path from $\mathcal{E}$}

\SetKwProg{myproc}{Procedure}{}{}
\myproc{\textsf{AllMaxSafe-BottomUp}$(\model,\mathcal{E})$}{
$\mathcal{L} = \{ P_i[l_i-1,r_i] : P_i[l_i,r_i] \in \mathcal{E}, l_i > 1\}$\
$\mathcal{R} = \{ P_i[l_i,r_i+1] : P_i[l_i,r_i] \in \mathcal{E}, r_i < |P_i|\}$\
$\mathcal{P} = \mathcal{L} \cup \mathcal{R}$\
$\mathcal{S}$ = \textsf{GetSafe}$(\model,\mathcal{P})$\
\For{$P_i[l_i,r_i] \in \mathcal{E}$}{
    \If{$P_i[l_i-1,r_i] \notin \mathcal{S}$ \kwAnd $P_i[l_i,r_i+1] \notin \mathcal{S}$}
    {
        \kwOutput $P_i[l_i,r_i]$\
    }
}
\If{$\mathcal{S}\ne \emptyset$}
    {
        \textsf{AllMaxSafe-BottomUp}$(\model,\mathcal{S})$\
    }
}

\caption{An algorithm to output all maximal safe subpaths that can be extended from an extending core set $\mathcal{E}$.\label{algo:allmax-bottomup}}
\end{algorithm}

\subsection{The two-pointer algorithm}
\label{sec:twopointer}

As we observed in \cref{sec:safety_test}, we can test whether a single path $P$ is safe using one ILP call.
We will assume that this test is encapsulated as a procedure \textsf{IsSafe}$(\model,P)$.
Once we can test whether a single path is safe for $\model(\vars,\cons)$, we can adopt a standard approach to compute all maximal safe paths. Namely, we start by computing one solution of $\model(\vars,\cons)$, $P_1,\dots,P_k$ and then compute maximal safe paths by a two-pointer technique that for each path $P_i$, finds all maximal safe paths by just a \emph{linear} number of calls to the procedure \textsf{IsSafe}~\cite{Khan:2022wo}.

\begin{figure}[tbh]
\centering
\begin{subfigure}[c]{0.32\textwidth}
\includegraphics[width=\textwidth]{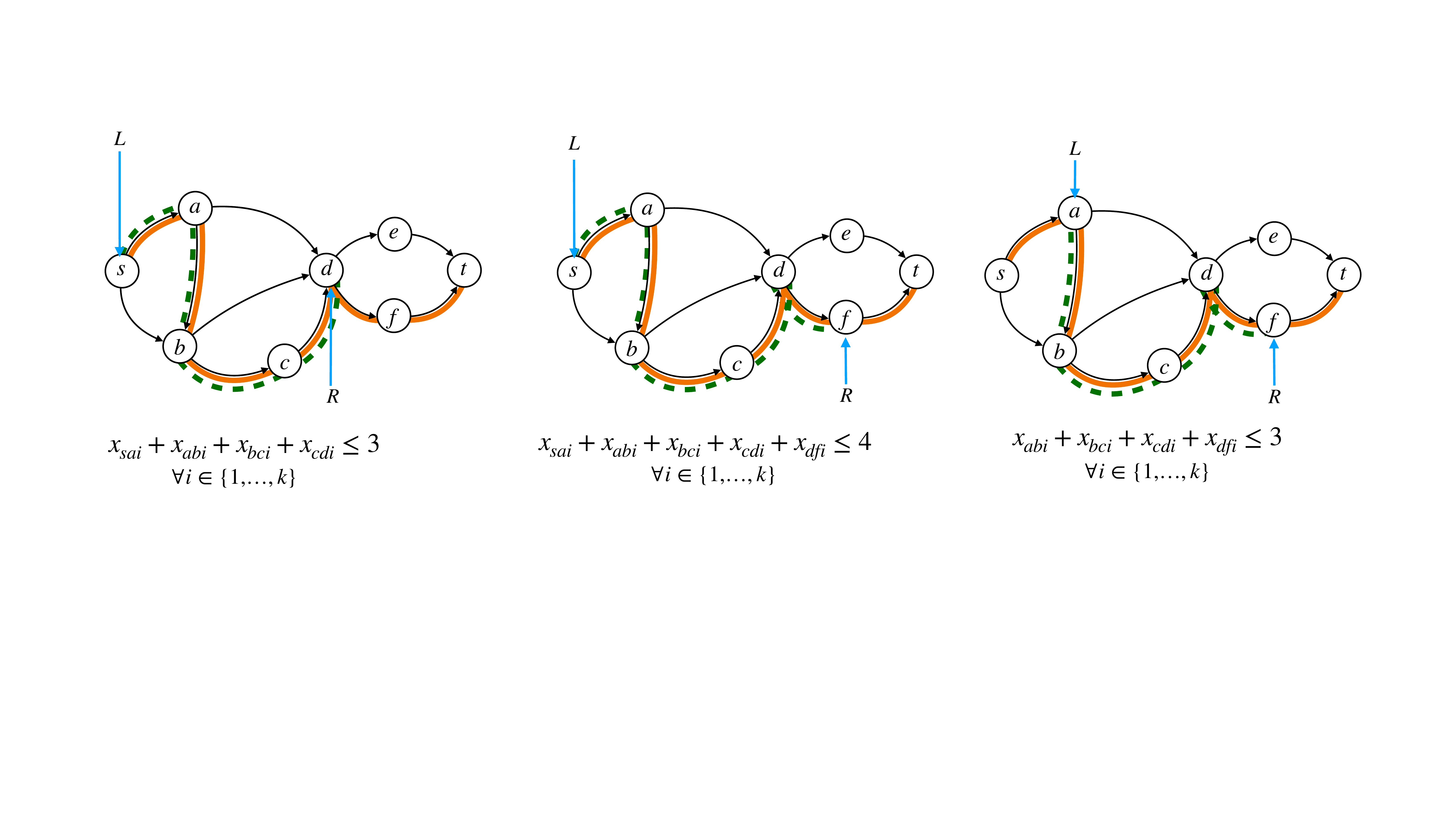}
\caption{Current iteration \label{fig:1-TF}}
\end{subfigure}
\begin{subfigure}[c]{0.33\textwidth}
\includegraphics[width=\textwidth]{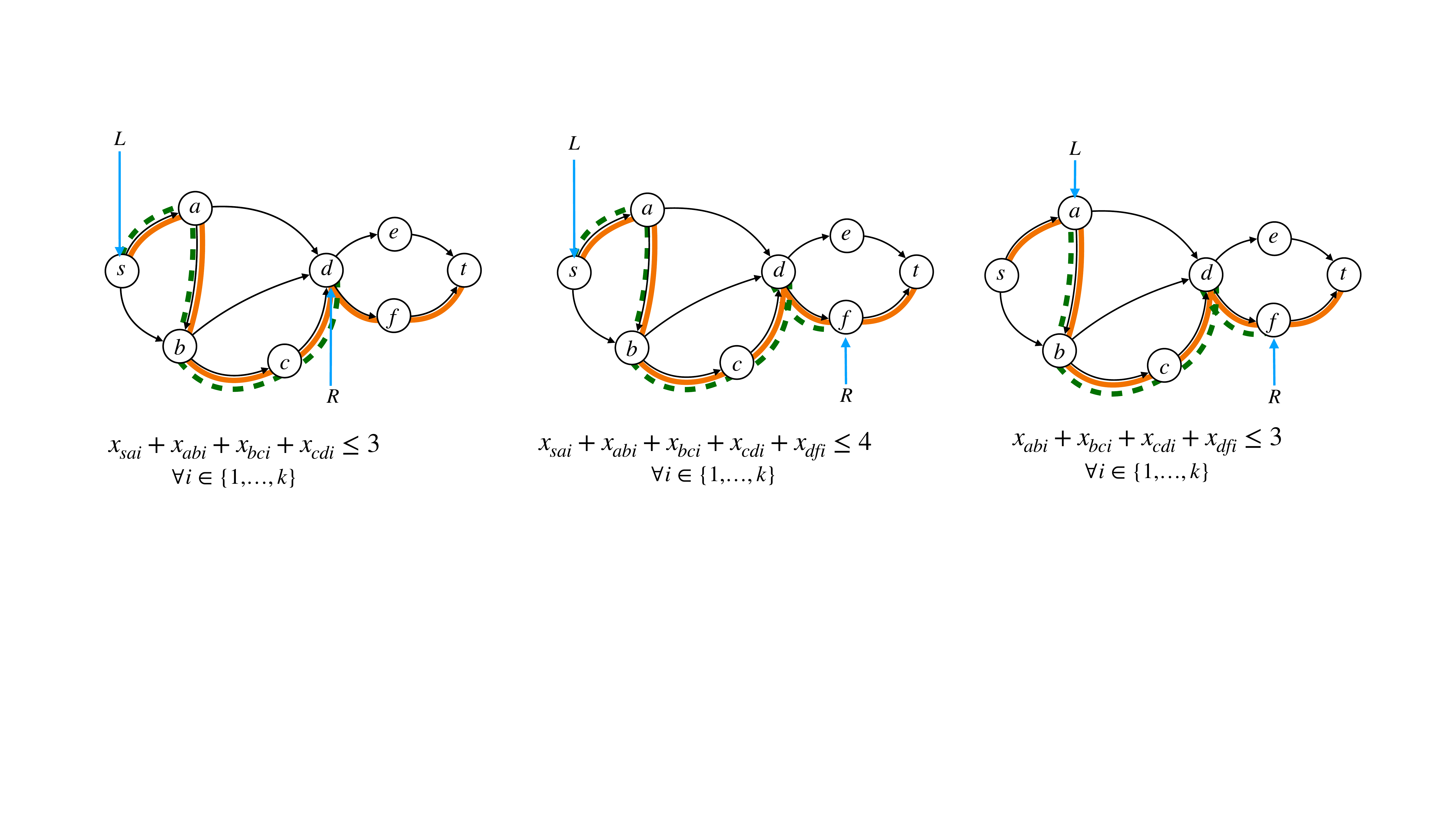}
\caption{Right pointer movement\label{fig:2-TF}}
\end{subfigure}
\begin{subfigure}[c]{0.33\textwidth}
\includegraphics[width=\textwidth]{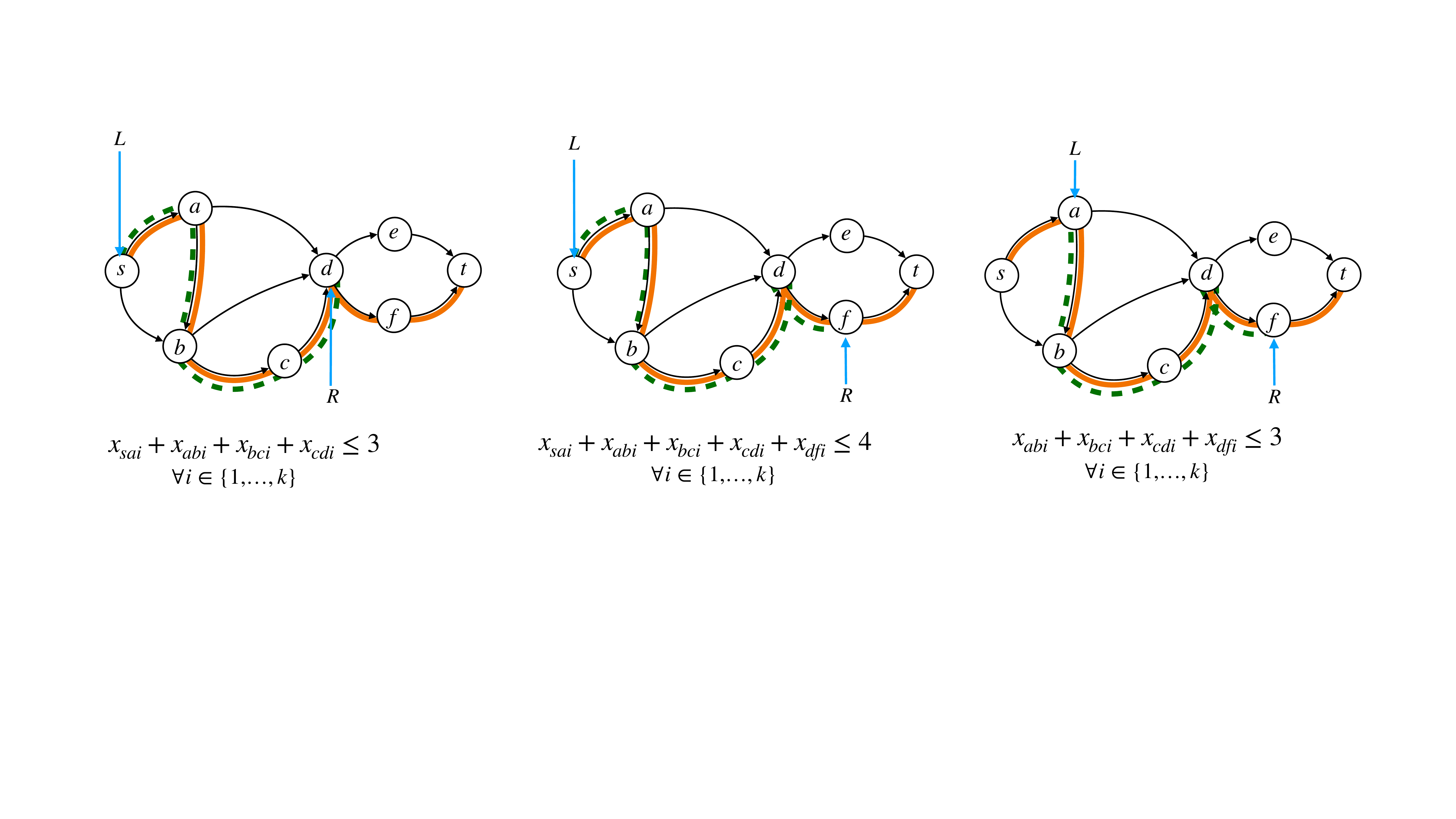}
\caption{Left pointer movement\label{fig:3-TF}}
\end{subfigure}
\caption{Illustration of the two-pointer algorithm applied on a flow decomposition path $P_i$ (in orange). In each sub-figure, the subpath $P$ (dashed green) between the nodes pointed by the left pointer $L$ and the right pointer $R$ is tested for safety, by adding constraints $\scon(P)$. In \subref{fig:1-TF}, \textsf{IsSafe}$(\model,P)$ returns \emph{True}, and the right pointer advances on $P_i$. In \subref{fig:2-TF}, \textsf{IsSafe}$(\model,P)$ returns \emph{False}, and the previous subpath from \subref{fig:1-TF} is output as a maximal safe path. In \subref{fig:3-TF}, the left pointer has advanced, and the new path $P$ is tested for safety.\label{fig:two-pointer-algorithms}}
\end{figure}

This works as follows. We use two pointers, a \emph{left} pointer $L$, and a \emph{right} pointer $R$. Initially, $L$ points to the first node of path $P_i$ and $R$ to the second node.
As long as the subpath of $P_i$ between $L$ and $R$ is safe, we move the right pointer to the next node on $P_i$. When this subpath is not safe, we output the subpath between $L$ and the previous location of $R$ as a maximal safe path, and we start moving the left pointer to the next node on $P_i$, until the subpath between $L$ and $R$ is safe. We stop the procedure once we reach the end of $P_i$. We summarize this procedure as \Cref{algo:allmax-twopointer}; see also \Cref{fig:two-pointer-algorithms} for an example.

\begin{algorithm}[hbt]
\allowdisplaybreaks
\SetKw{kwContinue}{continue}
\SetKw{kwReturn}{return}
\SetKw{kwOutput}{output}
\SetKw{kwAnd}{and}
\SetKw{kwNot}{not}
\SetKw{kwInfeasible}{infeasible}
\SetKw{kwFeasible}{feasible}
\SetKwRepeat{Do}{repeat}{until}
\KwIn{An ILP model $\model$ and one of its $k$ solution paths, $P_i = (v_1,\dots,v_t)$, $t \geq 2$}
\KwOut{All maximal safe subpaths of $P_i$ for $\model$}
\SetKwProg{myproc}{Procedure}{}{}
\myproc{\textsf{AllMaxSafe-TwoPointer}$(\model,P_i)$}{
$L \leftarrow 1$, $R \leftarrow 2$\
\While{True}{
    \While{\textsf{IsSafe}$(\model,P_i[L,R])$ \kwAnd $R \leq t$}
    {
        $R \leftarrow R + 1$\
    }
    \kwOutput $P_i[L,R-1]$\
    \lIf{$R > t$}
    {
        \kwReturn
    }
    \While{\kwNot \textsf{IsSafe}$(\model,P_i[L,R])$}
    {
        $L \leftarrow L + 1$\
    }
}
}

\caption{The two-pointer algorithm applied to compute all maximal subpaths of a given solution path $P_i$\label{algo:allmax-twopointer}}
\end{algorithm}

\subsection{Running time experiments among different variants proposed}\label{sec:variants-comparison}

We conducted the experiments on an isolated Linux server with AMD Ryzen Threadripper PRO 3975WX CPU with 32 cores (64 virtual) and 504GB of RAM. Time and peak memory usage of each program were measured with the GNU \texttt{time} command. \emph{SafeMFD} was allowed to run \texttt{Gurobi} with 12 threads. All C++ implementations were compiled with optimization level 3 (\emph{-O3} flag). Running time and peak memory is computed and reported per dataset.

\emph{SafeMFD} includes the following four variants computing maximal safe paths:

\begin{description}
\item[\emph{TopDown}]: Implements \Cref{algo:allmax-topdown} using the group testing in \Cref{algo:pathset-safety}.
\item[\emph{BottomUp}]: Implements \Cref{algo:allmax-bottomup} (\cref{sec:bottomup}) using the group testing in \Cref{algo:pathset-safety}.
\item[\emph{TwoPointer}]: Implements \Cref{algo:allmax-twopointer} (\cref{sec:twopointer}), the traditional two-pointer algorithm~\cite{Khan:2022wo}.
\item[\emph{TwoPointerBin}]: Same as previous variant, but it additionally replaces the \emph{linear scan} employed to extend and reduce the currently processed safe path by a \emph{binary search}\footnote{The binary search is only applied if the search space is larger than a constant threshold set experimentally.}.
\end{description}

To compare between our four different variants we first run them all on every dataset, and then filter out those graphs that ran out of time in some variant. This way we ensure that no variant consumes its time budget and thus our running time measurements are not skewed by the unsuccessful inputs' timeouts. Applying this filter we removed 83 graphs from the \emph{Catfish} dataset (0.3\%) and 4,515 graphs from the \emph{RefSim} dataset (43.74\%).

\begin{table}[t]
\centering
\begin{tabular}{|c|c|c|c|c|}
\hline
Dataset (\# Graphs) & Variant & Time (hh:mm:ss) &  \# ILP calls\\ \hline
\multirow{3}{*}{\parbox[c]{2cm}{\begin{center}Catfish\\(27,613)\end{center}}}
& TopDown & 01:13:27 & 124,676\\
& BottomUp & 03:22:13 & 212,774\\
& TwoPointer & 04:21:44 & 226,365\\
& TwoPointerBin & 03:31:57 & 216,540\\ \hline
\multirow{3}{*}{\parbox[c]{2cm}{\begin{center}RefSim\\(5,808)\end{center}}}
& TopDown & 04:38:41 & 55,450\\
& BottomUp & 11:55:20 & 76,837\\
& TwoPointer & 13:48:00 & 127,352\\
& TwoPointerBin & 11:34:02 & 119,218\\ \hline
\end{tabular}
\caption{Running times and number of ILP calls in four different variants of \emph{SafeMFD}.}
\label{tab:performance_variants}
\end{table}

\Cref{tab:performance_variants} shows the running times and number of ILP calls of the different variants on both datasets. \emph{TopDown} clearly outperforms the rest, being at least twice as fast, and performing (roughly) half many ILP calls. While \emph{BottomUp} is analogous to \emph{TopDown}, the superiority of the latter can be explained by the length maximal safe paths. Since maximal safe paths are long it is faster to obtain them by reducing unsafe paths (\emph{TopDown}) than by extending safe paths (\emph{BottomUp} and both \emph{TwoPointer} variants). On the other hand, \emph{TwoPointer} is the slowest variant and \emph{BottomUp} and \emph{TwoPointerBin} obtain similar improvements (over \emph{TwoPointer}) by following different strategies. While \emph{BottomUp} reduces the number of ILP calls more than \emph{TwoPointerBin} (better appreciated in the RefSim dataset), the ILP calls of \emph{BottomUp} take longer (since \emph{BottomUp} tests several paths at the same time and \emph{TwoPointerBin} only one), and thus the total running times of both is similar. This motivates future work on combining both approaches, while processing the paths starting from unsafe (as in \emph{TopDown}) for better performance.





\section{Hardness of Testing MFD Safety}
\label{sec:hardness}

In this section we give a Turing-reduction from the UNIQUE 3SAT problem (U3SAT) to the problem of determining if a given path $P$ in
a flow network $G$ is safe for minimum flow decomposition (call this problem \emph{MFD-SAFETY}).  A 3SAT instance belongs to U3SAT if and only if it has exactly
one satisfying assignment.
U3SAT has been shown to be NP-hard under randomized reductions~\cite{VALIANT198685}, but it is open as to whether it is
NP-hard in general.

The reduction leverages the construction in \cite{hartman2012split} that reduces 3SAT to minimum flow decomposition.
We first briefly review this construction:
A variable gadget (see Fig.~4 in \cite{hartman2012split}) is created for each 3SAT variable $x$ and
a clause gadget (see Fig.~5 in \cite{hartman2012split}) is created for each 3SAT clause.  Positive literals in
each clause receive flow from the left side of the corresponding variable gadget, whereas negative literals receive
flow from the right side.  Theorem VI.1 in \cite{hartman2012split} establishes that a 3SAT instance is satisfiable if and only if  the constructed flow network has a minimum flow decomposition of a certain size.  Any flow decomposition achieving this size must have a specific structure; in particular, there must be
a flow path of weight $4$ that either travels up the left side of the gadget (setting $x$ to TRUE), or the right side (setting $x$ to FALSE).

\begin{figure}[ht]
\centering
\includegraphics[width=0.16\textwidth]{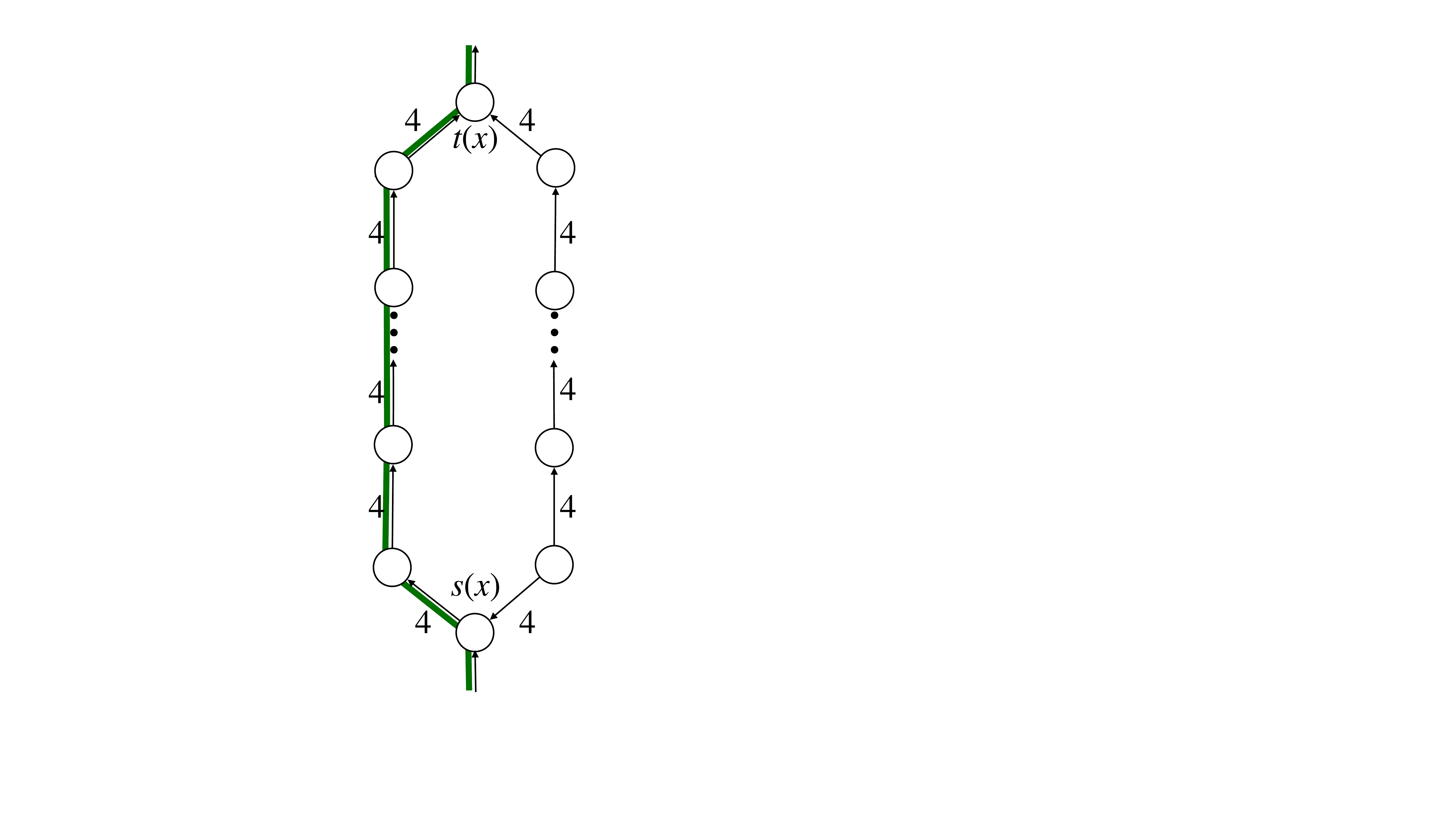}
\caption{The variable gadget from \cite{hartman2012split}, showing only the weight $4$ edges (other edges have weights from $\{1,2\})$.
A key property established in \cite{hartman2012split} is that if the 3SAT instance is satisfiable then in a minimum
flow decomposition, a weight $4$ flow path must travel up either the left side of the gadget (as shown), or the right side.
A left flow path indicates the variable should be set to TRUE, while right indicates FALSE.
We leverage this construction to reduce U3SAT to MFD-SAFETY.}
\end{figure}

\begin{theorem}
There is a polynomial time Turing-reduction from U3SAT to MFD-SAFETY.
\end{theorem}

\begin{proof}
To obtain the desired Turing-reduction algorithm, instead of checking the size of
the MFD, we will instead sequentially check the MFD-SAFETY of
the aforementioned \emph{side paths} traveling up the left and right sides of each variable gadget.
Provided each variable gadget has exactly one safe
side path we can then check the corresponding truth assignment to see if each clause is satisfied.
If yes, we accept the instance as belonging to U3SAT, otherwise we reject.

Suppose the instance does belong to U3SAT.  In this case there is a satisfying assignment so the MFD must
have the structure as described above.  Furthermore, since there is exactly
one satisfying assignment, exactly one side path of each variable gadget must be safe and so our algorithm finds it
and then verifies that the truth assignment satisfies each clause, thus accepting the instance.
On the other hand, if the instance does not belong to U3SAT it could either be unsatisfiable or have multiple
satisfying assignments.  If unsatisfiable, no matter whether the safety  checks pass, the corresponding assignment will
not satisfy all clauses, so the instance will be rejected.  If there are multiple solutions, then any
variable that can be both TRUE and FALSE will not have a safe side path in the MFD.
This means the safety check will fail and the instance will again be rejected.
\qed
\end{proof}

\end{document}